\documentclass[10pt,journal,twocolumn,twoside]{IEEEtran}

\usepackage{amsmath,amsfonts}
\usepackage{algorithmic}
\usepackage{algorithm}
\usepackage{array}
\usepackage[caption=false,font=normalsize,labelfont=sf,textfont=sf]{subfig}
\usepackage{textcomp}
\usepackage{stfloats}
\usepackage{url}
\usepackage{verbatim}
\usepackage{graphicx}
\usepackage{cite}
\usepackage{color}
\usepackage{setspace}
\newtheorem{thm}{\bf Theorem}
\newtheorem{rem}{Remark}

\renewcommand{\IEEEQED}{\IEEEQEDopen}

\begin{document}

\title{Wideband Channel Sensing with Holographic Interference Surfaces}

\author{Jindiao Huang, and Haifan Yin, ~\IEEEmembership{Senior Member,~IEEE}
        
\thanks{This work was supported by the Major Program (JD) of Hubei Province (2025BEA001), the Fundamental Research Funds for the Central Universities and the National Natural Science Foundation of China under Grant 62071191. \textit{(Corresponding author: Haifan Yin.)}}
\thanks{The authors are with the School of Electronic Information and Communications, Huazhong University of Science and Technology, Wuhan 430074, China (e-mail: jindiaohuang@hust.edu.cn; yin@hust.edu.cn)}}



\maketitle

\begin{abstract}
The Holographic Interference Surface (HIS) opens up a new prospect for building a more cost-effective wireless communication architecture by performing Radio Frequency (RF) domain signal processing. In this paper, we establish a wideband channel sensing architecture for electromagnetic wave reception and channel estimation based on the principle of holographic interference theory. Dute to the nonlinear structure of holograms, interferential fringes composed of wideband RF signals exhibit severe self-interference effects in the time-frequency domain, which are inherently resistant to the classical signal processing tools. To overcome the self-interference, we propose a holographic channel recovery method, which analyzes the time-domain variation of holograms from a geometrical perspective and constructs an inverse mapping from wideband holograms to object waves. Based on the Wirtinger partial derivative and Armijo condition, we then develop a wideband hologram-based maximum likelihood (WH-ML) estimation method for estimating the channel state information (CSI) from holograms. We also propose a geometric rotation-based object wave sensing (GROWS) algorithm to address the complicated computation of ML estimation. Furthermore, we derive the Cram\'er-Rao lower bound (CRLB) for investigating the achievable performance of wideband holographic channel estimation. Simulation results show that under the wideband channel sensing architecture, our proposed algorithm can accurately estimate the CSI in wideband scenarios.
\end{abstract}

\begin{IEEEkeywords}
Holographic interference surfaces, holographic communication, wideband channel sensing, CSI, holographic interference principle.
\end{IEEEkeywords}

\section{Introduction}
\IEEEPARstart{A}{long} with the digital development of society and economy, the sixth generation (6G) communication system is expected to be a ubiquitous distributed intelligent network, realizing the organic integration of communication, perception and computation, providing a communication foundation for the era of ``Internet of Everything" \cite{Chen:23JSAC, Chettri:20ITJ, Saad:20Network}. To fulfill the requirement of scenarios such as extended reality, smart cities and intelligent factories, 6G is expected to achieve Tbps-scale data rates, Kbps/Hz-scale spectral efficiency and $\mu$s-level latency \cite{You:23WC}. Benefiting from the development of massive multiple-input multiple-output (MIMO) and beamforming technologies, 5G communications substantially improve the communication rates and reliability on the basis of 4G \cite{Marzetta:10TWC}. Although large-scale phased arrays with hundreds or more antennas show promising potential for 6G communications, the ensuing issues of increased hardware complexity, escalated power consumption and expanded array size tremendously limit their deployment in practical scenarios \cite{Emil:17FTSP}. 

To break through the limitations of massive MIMO, researchers are diligently exploring novel communication architectures with enhanced cost-effectiveness, in which an immensely promising technology is holographic MIMO (HMIMO) \cite{Gong:24CST, An:23CL, Dardari:20JSAC}. As an extension to massive MIMO, HMIMO realizes dense or even quasi-continuous arrangements of antennas in a limited space by reducing the antenna spacing to smaller than half a wavelength \cite{Pizzo:20JSAC}. An introduction of HMIMO is presented in \cite{Huang:20WC}, in which it provides a general overview of the hardware architecture, characteristics and application scenarios of HMIMO. In \cite{Pizzo:20}, the channel spatial degrees of freedom (DoF) are obtained utilizing Fourier plane-waves series expansion, exposing the optimal deployment of HMIMO antenna units. In order to investigate the performance enhancement of HMIMO by triple polarization, the work of \cite{Wei:23TWC} constructs a near-field channel model capable of describing the characteristics of triple polarization in multi-user communication scenarios. The authors of \cite{An:23JSAC} present a stacked metasurface layers-enabled HMIMO communication system that is capable of performing precoding with reduced number of RF chains. In \cite{Ji:24TWC}, an electromagnetic hybrid beamforming scheme for HMIMO communication is proposed for eliminating the inherent radiation pattern distortion of commercial base station antenna arrays. In HMIMO architecture, the Fresnel region increases significantly due to the quasi-continuous arrangement of the antenna units, which leads to hybrid near-far field communication. To estimate the channel state information (CSI) under such a hybrid-field case, the authors of \cite{Yue:24TWC} developed a power-diffusion-aware hybrid-field channel estimation method.

Currently, most studies related to holographic communication focus primarily on the dense deployment of antenna elements, exploring the performance gains and wireless signal transceiving challenges associated with compressed antenna spacing. Although densely placed antenna arrays show promise for deploying ultra-large-scale arrays in limited spaces, they are fundamentally unable to address issues of increased power consumption and hardware complexity, which stem from the reliance of existing signal processing on radio frequency (RF) chains. Nevertheless, only a few studies have focused on exploiting holographic principles to enhance signal processing architectures. Therefore, this paper concentrates on exploring approaches to realize RF-domain signal processing through RF signal interference with the aim of achieving simplified hardware design and improved energy efficiency.
It is essential to highlight that our research is fundamentally different from HMIMO in terms of motivation and content. HMIMO capitalizes on the coupling effect of denser arrays to minimize sidelobe energy leakage and enhance array gain. In contrast, our research leverages the interference effect of signals for channel estimation, thereby reducing the number of RF chains and improving the system energy efficiency.

In our previous work \cite{Huang:24TWC}, we developed the holographic interference surface (HIS), a metamaterial-based antenna array detecting wireless signals based on the interferometric superposition of RF signals. Different from the hardware architecture of conventional receivers, HIS avoids complex operations such as filtering, analog-to-digital conversion (ADC), and down-conversion, requiring only envelope detectors and an RF chain for channel estimation, which provides a novel paradigm for the construction of a more cost-effective wireless communication architecture. Based on HIS, we construct a holographic channel estimation architecture for narrowband signals in \cite{Huang:24TWC}. For narrowband scenarios, the hologram only behaves as a quadratic function of the channel matrix and it does not exhibit time-varying properties for stationary users. However, in wideband scenarios, the nonlinear characteristics of the holograms are dramatically magnified, which leads to severe coupling between the components of the channel matrix at different frequencies. Therefore, new holographic channel estimation methods are required to deal with the interference caused by the nonlinear characteristics of holograms.

In this paper, we construct the architecture of wideband holographic communication system based on the holographic interference principle. In order to address the spectral overlapping caused by the nonlinear characteristics of holograms, we analyze the structure of wideband holograms and propose a holographic channel recovery method. We then analyze the probability distribution properties of holograms and develop a wideband hologram-based maximum likelihood (WH-ML) estimation algorithm for estimating CSI from holograms. To avoid the complicated computation of ML estimation, we also proposed a low complexity alternative, called geometric rotation-based object wave sensing (GROWS) algorithm. Furthermore, the Cram\'er-Rao lower bound (CRLB) of the wideband holographic channel estimation problem is derived.

The main contributions of our paper are as follows:
\begin{itemize}
    \item{We propose a wideband channel sensing architecture with HISs. The principles of holographic interference theory for wideband RF signal reception and channel estimation are derived.}
    \item{Based on a theoretical analysis of the probability distribution characteristics of wideband holograms, we develop a maximum likelihood estimation to extract the channel state information from wideband holograms.}
    \item{We present the holographic channel recovery method to reduce the self-interference caused by the nonlinear features of the hologram, based on which the GROWS algorithm is proposed as a generalized holographic channel estimation method with much lower complexity.}
    \item{We derive the CRLB of the wideband holographic channel estimation problem. Our numerical results demonstrate that our proposed channel sensing architecture effectively solves the holographic channel estimation problem with a highly simplified hardware design.}
\end{itemize}

The rest of this paper is organized as follows. In Sec. \ref{secSysMdl}, the uniform planar array (UPA) channel model of 3GPP is introduced \cite{3GPP:TR38901}. In Sec. \ref{secHoloComm} we first derive the basic principles of signal receiving for wideband holographic communication systems, then propose the holographic channel recovery method. The maximum likelihood estimation, the GROWS algorithm and the CRLB are derived in Sec. \ref{secHCE}. Sec. \ref{secNumRes} contains the simulation results and the final conclusions are drawn in Sec. \ref{secConcl}.

\textit{Notation:} Matrices and vectors are denoted by boldface letters. For a matrix $\mathbf{X}$, its inverse, conjugate, transpose and conjugate transpose are denoted by $\mathbf{X}^{-1}$, $(\mathbf{X})^{*}$, $(\mathbf{X})^{T}$ and $(\mathbf{X})^{H}$ respectively. The Kronecker product of $\mathbf{X}$ and $\mathbf{Y}$ is denoted by $\mathbf{X}\otimes\mathbf{Y}$. The $ \ell_2 $ norm of a vector or the spectral norm of a matrix is denoted by the same symbol $\|\cdot\|_2$. diag$\{\mathbf{a_1},\cdots,\mathbf{a_N}\}$ is a diagonal matrix with $\mathbf{a_1},\cdots,\mathbf{a_N}$ at the main diagonal. $\mathbb{C}$ denotes the set of complex numbers. $|\cdot|$ and arg($\cdot$) stand for the amplitude and the phase of a complex number respectively. The real part and imaginary part of a complex number are denoted as $\mathfrak{R}(\cdot)$ and $\mathfrak{I}(\cdot)$ respectively. $\mathcal{CN}(\mu,\sigma^2)$ represents the complex univariate Gaussian distribution with the mean $\mu$ and variance $\sigma^2$. $\mathcal{X}_k^2(\lambda,\sigma^2)$ denotes the non-central chi-squared distribution with degrees of freedom $k$, centrality parameter $\lambda$ and variance parameter $\sigma^2$. The degree-$k$ non-central chi distribution with non-centrality parameter $\lambda$ is denoted by $\mathcal{X}_k(\lambda)$. $\mathbb{E}[\cdot]$ stands for the expectation and $\text{Cov}(\cdot)$ denotes the covariance matrix. $\overset{\Delta}{=}$ is used for definition.

\section{System Model} \label{secSysMdl}
\begin{figure}[!t]
    \centering
    \includegraphics[width=\columnwidth]{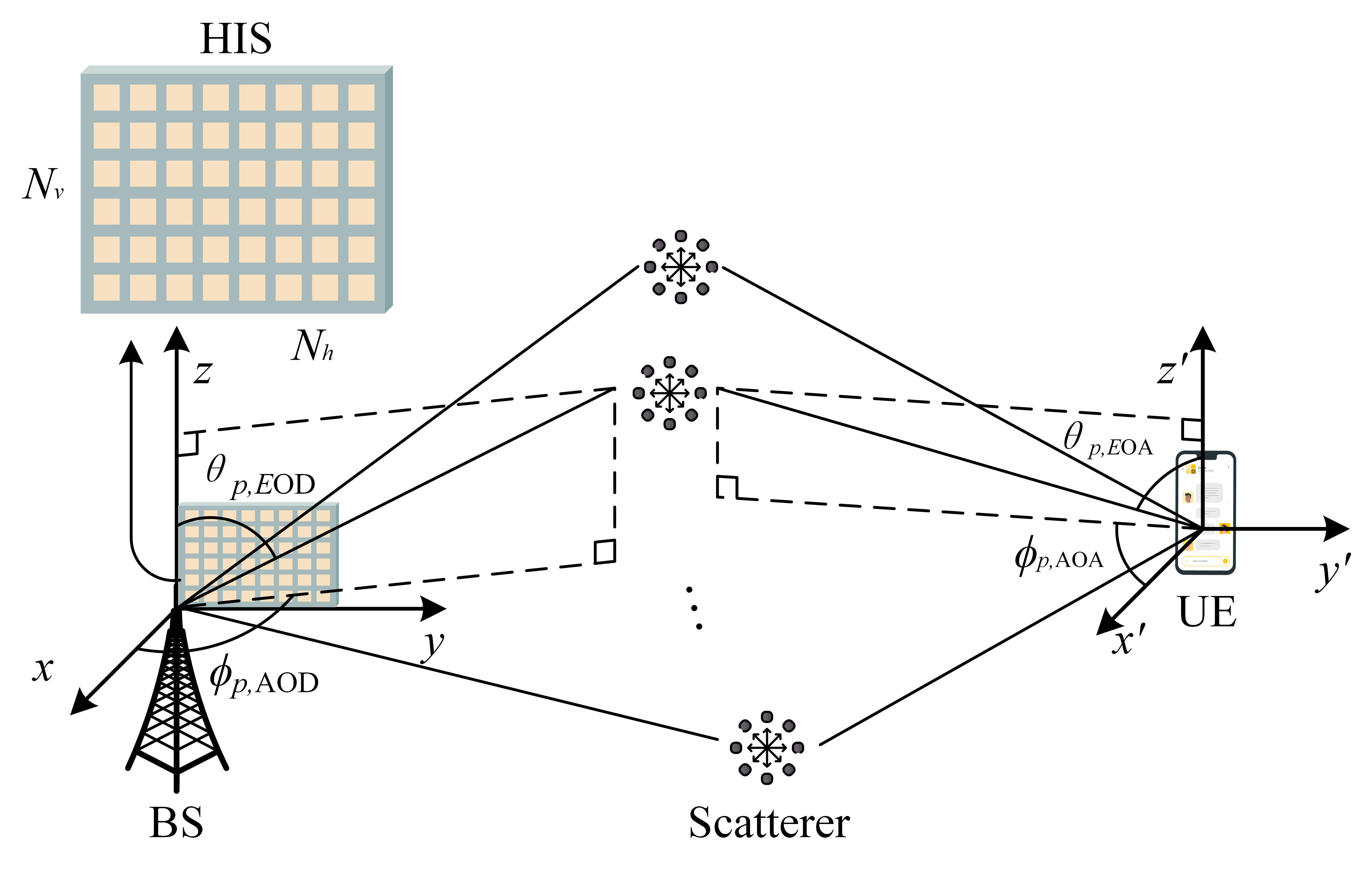}
    \caption{The channel between the BS and the UE.}
    \label{figSysMdl}
\end{figure}

We consider a wideband system as shown in Fig. \ref{figSysMdl}, where the base station (BS) is equipped with the HIS composed of $N_v$ rows and $N_h$ columns of radiation units. The horizontal and vertical unit spacing are denoted by $D_h$ and $D_v$. The number of the HIS units is $N_t = N_v N_h$ and the number of antennas at the user equipment (UE) is 1. The channel contains multiple clusters with each consisting of several scattering rays, constituting a total of $P$ multipath.

Denote the channel between the HIS and a certain UE at frequency $f$ and time $t$ by $\mathbf{h}(f,t)$, which is expressed as \cite{3GPP:TR38901}
\begin{align}
    \mathbf{h}(f,t) 
        \!=\!\! \sum_{p=1}^P \!\beta_p e^{-j2\pi f \tau_p} e^{j \frac{2\pi \hat{\mathbf{r}}_{rx,p}^T \overline{\mathbf{d}}_{rx}}{\lambda_0}} \! e^{j 2\pi \omega_p t} \! \mathbf{a}(\theta_p, \phi_p), \label{eqhft}
\end{align}
where $\beta_p$ and $\tau_p$ denote the complex amplitude and the delay of the $p$-th path respectively. The wavelength $\lambda_0$ is given by the central carrier frequency $f_c$ and the speed of light $c$ as $\lambda_0 = c/f_c$. Additionally, the spherical unit vector of the $p$-th path $\hat{\mathbf{r}}_{rx,p}$ is defined as
\begin{align}
    \hat{\mathbf{r}}_{rx,p} = 
        \begin{bmatrix}
            \sin{\theta_{p,\text{ZOA}}} \cos{\phi_{p,\text{AOA}}} \\
            \sin{\theta_{p,\text{ZOA}}} \sin{\phi_{p,\text{AOA}}} \\
            \cos{\phi_{p,\text{ZOA}}}
        \end{bmatrix},
\end{align}
where $\theta_{p,\text{ZOA}}$ and $\phi_{p,\text{AOA}}$ represent the zenith angle of arrival (ZOA) and azimuth angle of arrival (AOA), respectively. Furthermore, $\overline{\mathbf{d}}_{rx}$ is the location vector of the UE. The Doppler of the $p$-th path is defined as $\omega_p = \hat{\mathbf{r}}_{rx,p}^T \mathbf{v}/\lambda_0$ with the velocity vector of the UE:
\begin{align}
    \mathbf{v} = v 
        \begin{bmatrix}
            \sin{\theta_v} \cos{\phi_v} & \sin{\theta_v} \sin{\phi_v} & \cos{\theta_v}
        \end{bmatrix}^T,
\end{align}
where the speed, velocity zenith angle and velocity azimuth angle are denoted by $v$, $\theta_v$ and $\phi_v$, respectively. The last term in Eq. (\ref{eqhft}) represents the 3-D steering vector of the $p$-th path given by
\begin{align}
    \mathbf{a}(\theta_p, \phi_p) = \mathbf{a}_h(\theta_p,\phi_p) \otimes \mathbf{a}_v(\theta_p,\phi_p) \in \mathbb{C}^{N_t},
\end{align}
where
\begin{align}
    \mathbf{a}_h(\theta,\phi)
        =&  \begin{bmatrix}
                1 \!\!&\!\! e^{j 2\pi \frac{D_h \sin{\theta} \sin{\phi}}{\lambda_0}} \!\!\!&\!\!\! \cdots \!\!&\!\! e^{j 2\pi \frac{(N_h \!-\! 1) \! D_h \! \sin{\theta} \sin{\phi}}{\lambda_0}}
            \end{bmatrix}^T \!\!\!,\\
    \mathbf{a}_v(\theta,\phi)
        =&  \begin{bmatrix}
                1 & e^{j 2\pi \frac{D_v \cos{\theta}}{\lambda_0}} & \cdots & e^{j 2\pi \frac{(N_v - 1) D_v \cos{\theta}}{\lambda_0}}
            \end{bmatrix}^T.
\end{align}

We consider a wideband system composed of $N_f$ subcarriers with spacing $\Delta f$ and the matrix form of the channel between the BS and a certain UE at all subcarriers is represented as
\begin{align}
    \mathbf{H}(t) = 
        \begin{bmatrix}
            \mathbf{h}(f_1,t) & \mathbf{h}(f_2,t) & \cdots & \mathbf{h}(f_{N_f},t) 
        \end{bmatrix}. \label{eqH}    
\end{align}
According to \cite{Yin:20JSAC}, $\mathbf{H}(t) \in \mathbb{C}^{N_t \times N_f}$ is rewritten based on the angular-frequency-domain structure as 
\begin{align}
    \mathbf{H}(t) = \mathbf{A} \mathbf{C}(t) \mathbf{B}, \label{eqHu}
\end{align}
where the 3-D steering matrix $\mathbf{A} \in \mathbb{C}^{N_t \times P}$ is given by
\begin{align}
    \mathbf{A} = 
        \begin{bmatrix}
            \mathbf{a}(\theta_{1,\text{ZOD}},\phi_{1,\text{AOD}})&\cdots&\mathbf{a}(\theta_{P,\text{ZOD}},\phi_{P,\text{AOD}})
        \end{bmatrix}.   
\end{align}
The diagonal matrix $\mathbf{C}(t) \in \mathbb{C}^{P \times P}$ contains the Doppler of all paths, which is expressed as
\begin{align}
    \mathbf{C}(t) = \text{diag} \left\{ c_{1}(t), \cdots, c_{P}(t) \right\},
\end{align}
where
\begin{align}
    c_{p}(t) = \beta_p e^{j \frac{2\pi\hat{r}_{rx,p}^T\overline{d}_{rx}}{\lambda_0}} e^{j\omega_pt}.
\end{align}
$\mathbf{B} \in \mathbb{C}^{P \times N_f}$ represents the delay response at all subcarriers:
\begin{align}
\mathbf{B} = 
        \begin{bmatrix}
            \mathbf{b}(\tau_1)&\mathbf{b}(\tau_2)&\cdots&\mathbf{b}(\tau_P)
        \end{bmatrix}^T,    
\end{align}
where the delay response vector $\mathbf{b}(\tau_p) \in \mathbb{C}^{N_f}$ is defined as
\begin{align}
    \mathbf{b}(\tau_p) = 
        \begin{bmatrix}
            e^{-j2\pi f_1 \tau_p}&e^{-j2\pi f_2 \tau_p}&\cdots&e^{-j2\pi f_{N_f} \tau_p}
        \end{bmatrix}^T.    
\end{align}
Accordingly, the signal arriving at the HIS with additive white Gaussian noise $\omega \sim \mathcal{CN}(0,\sigma_{\omega}^2)$ can be written as
\begin{align}
    \mathbf{y} = \mathbf{H}(t) \mathbf{s} + \omega,
\end{align}
where the transmitted symbol of UE is denoted by $\mathbf{s}$.

In conventional systems, channel estimation is accomplished at the baseband, implicating complex operations such as filtering, analog-to-digital conversion, and down-conversion. However, HIS employs a fundamentally different signal processing architecture that performs channel estimation at the radio frequency and avoids the complicated operations mentioned above. The HIS enables interferometric superposition of the self-generated reference wave and the UE-transmitted signal through the combiner. Then, the power of the superimposed signal is captured as holograms utilizing power detection devices such as envelope detectors. Since the phase difference between the RF signal and reference wave is implicitly covered in holograms, it is possible to estimate the CSI by only dealing with the power of the received signal.

\section{Wideband Holographic Communication} \label{secHoloComm}
The holographic interference principle utilizes the interference phenomenon of electromagnetic waves to extract RF signals by changing the phase of reference waves. In this section, we first briefly introduce the holographic interference principle and analyze how to estimate the CSI based on the principle in Sec. \ref{secHoloComm_PHC}. Then, we extend the principle to wideband systems and propose a holographic channel recovery method to eliminate the self-interference of wideband holograms in Sec. \ref{secHoloComm_HCR}.

\subsection{Principle of Holographic Communication} \label{secHoloComm_PHC}
In our previous work \cite{Huang:24TWC}, we derived the principle of holographic interference theory for electromagnetic wave reception and transmission, exploring a novel scheme for channel estimation by only dealing with the power of the received signal. In the following, it is shown that signal receiving and channel estimation can also be performed for wideband signals based on the geometric characteristics of holograms.

\begin{figure}[!t]
    \centering
    \subfloat[]{\includegraphics[width=.9\columnwidth]{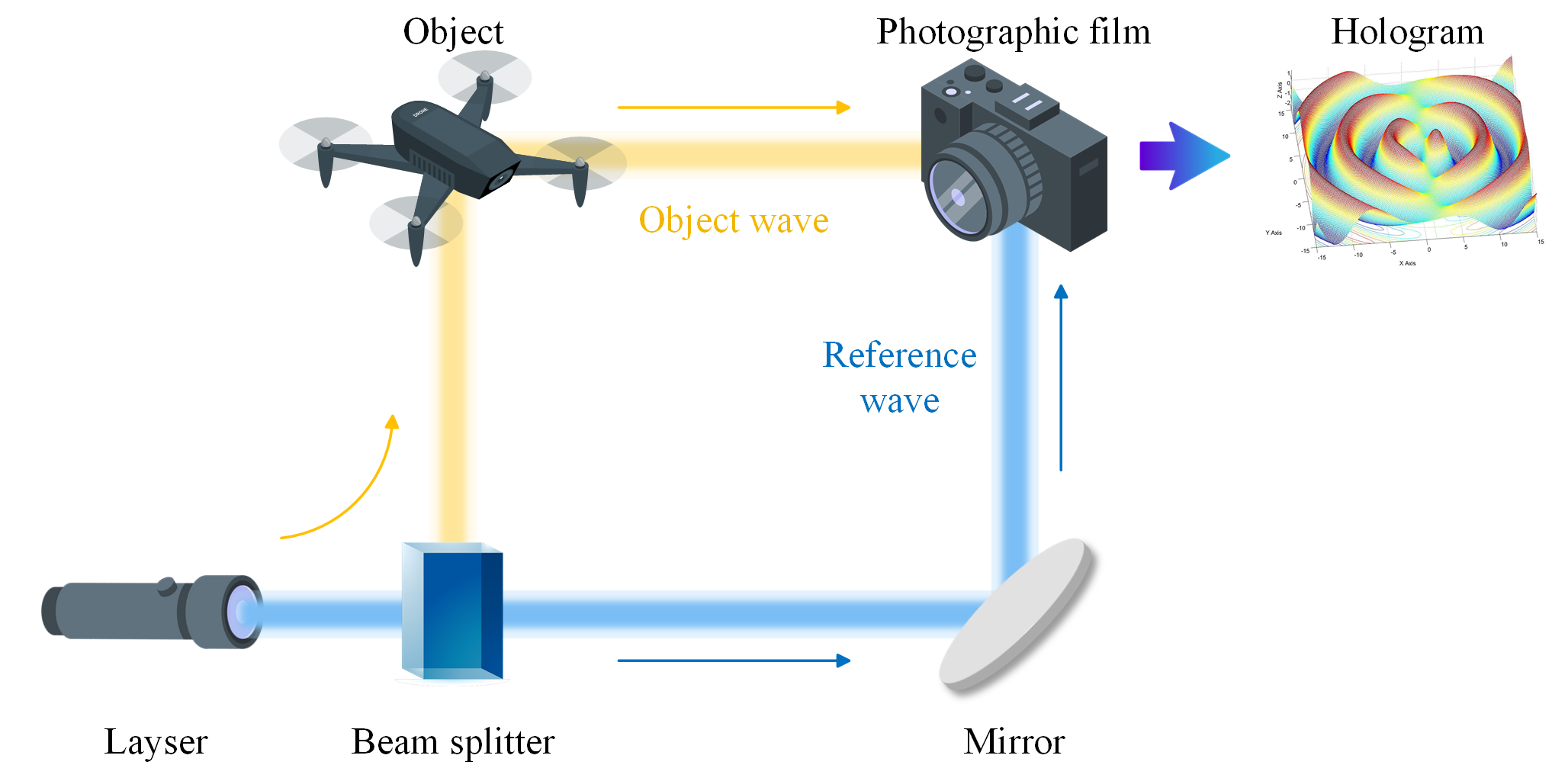} \label{figOptHolo}}\\
    \subfloat[]{\includegraphics[width=.9\columnwidth]{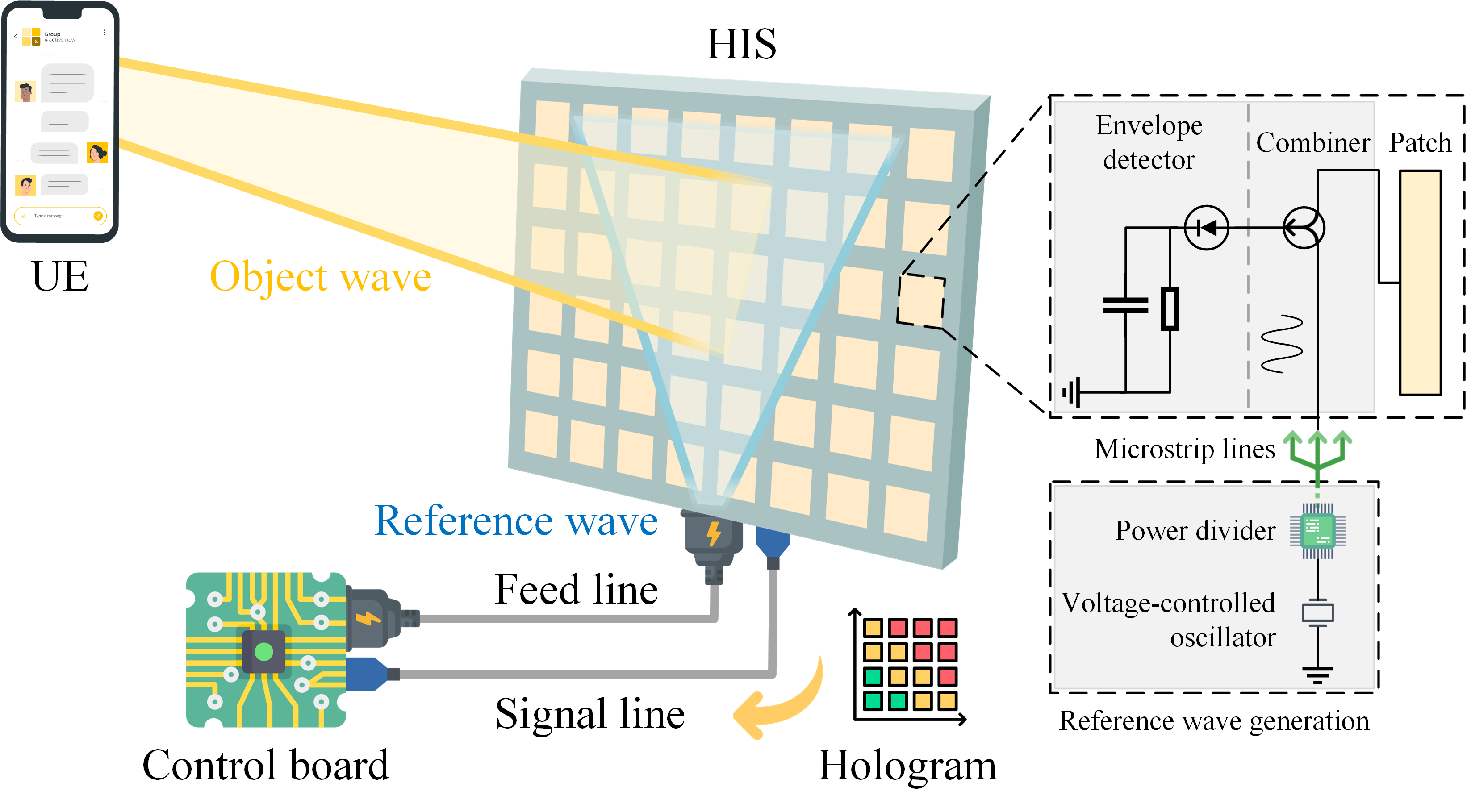} \label{figComHolo}}
    \caption{Illustration of holographic recording. (a) Optical holography. (b) Communication holography.}
    \label{figHolography}
\end{figure}

In the recording process of optical holography shown in Fig. \ref{figHolography}\subref{figOptHolo}, one light beam generated by a laser is split into two parts by a beam splitter. One part illuminates the object and reflects to the photographic film, called the object wave, and the other part propagates directly to the film via the mirror, called the reference wave. These two beams produce an interference superposition on the film, and this interference pattern is recorded as a hologram. In order to keep the consistency of terms in optical and communication holography, we denote the signal transmitted by the UE as the object wave, and the sinusoidal signal generated by HIS as the reference wave.

The basic steps for signal receiving based on the holographic principle include holographic interference, hologram recording and channel recovery. The HIS detects or estimates wireless signals based on the interferometric superposition of the object wave transmitted by UE and the reference wave generated by the surface. The intensity of the superimposed signal is recorded as the RF hologram by envelope detectors. Eventually, the phase of the received signal is recovered using multiple holograms to obtain the complex domain representation of the received signal, where the complex domain refers to the representation of a signal in the complex plane.

The hardware structure of the HIS is depicted in Fig. \ref{figHolography} \subref{figComHolo}. Initially, a sinusoidal reference signal is generated by a voltage-controlled oscillator. This signal is then transmitted to each unit via microstrip lines after passing through integrated power divider chips. Subsequently, the signal transmitted by the UE is superimposed with the reference signal using combiners. The hologram is then generated by measuring the power of the superimposed signal through envelope detectors. Finally, the CSI is estimated based on multiple holograms.

We consider a wideband OFDM system, thus the frequency of the $k$-th subcarrier satisfies $f_k = f_c + \frac{k-1}{T_s}$, where $f_c$ is the carrier frequency and $T_s$ is the symbol period. Based on the channel model introduced in Sec. \ref{secSysMdl}, the RF signal arriving at the $(m,n)$-th unit of HIS, i.e. the object wave, is given by:
\begin{align}
    E_o(t) 
    &= \mathbf{H}^{(m,n)}(t)\mathbf{s} \nonumber \\
    &= \sum_{k=1}^{N_f} \left( \sum_{p=1}^P a_p^{(m,n)} c_{p}(t) e^{j2\pi f_k (t - \tau_k)} \right),
\end{align}
where $\mathbf{H}^{(m,n)}(t)$ is the $(m N_v + n)$-th row of $\mathbf{H}(t)$ and
\begin{align}
    a_p^{(m,n)} = e^{j2\pi \left( m\frac{D_h \sin{\theta_p} \sin{\phi_p}}{\lambda_0} + n\frac{D_v\cos{\theta_p}}{\lambda_0} \right)}.
\end{align}

The subsequent signal processing is accomplished within two OFDM symbol periods, which is typically shorter than a coherence time. Therefore, we may further simplify $E_o(t)$ as:
\begin{align}
    E_o(t) = \sum_{k=1}^{N_f} h_k^{(m,n)} e^{j2\pi f_k t}, \label{eqEo}
\end{align}
where $h_k^{(m,n)}$ represents the CSI of the $(m,n)$-th unit of the $k$-th subcarrier and is given by
\begin{align}
    h_k^{(m,n)} = \sum_{p=1}^P a_p^{(m,n)} \beta_p e^{j2\pi \frac{\hat{r}_{rx,p}^T\overline{d}_{rx}}{\lambda_0}} e^{-j2\pi f_k \tau_p}.
\end{align}

The reference wave generated by the surface is given by $E_r(t) = A_r e^{j2\pi f_r t}$ with $A_r$ and $f_r$ denoting the amplitude and frequency of the reference wave, respectively. Suppose the object wave and reference wave interfere upon the HIS simultaneously and the envelope of the superimposed signal is recorded as the hologram. Thus, the hologram at the $(m,n)$-th unit of HIS is given by
\begin{align}    
    E_{\!I\!}(t) \!\!
        &= \!\! |E_r(t) + E_o(t)|^2 \label{eqI1}\\
        &= \!\! A_r^2 \!\!+\!\!\! \sum_{l=1}^{N_f} \!\! \sum_{k=1}^{N_f} \! |\!h_l^{(m,n)}\!| |\!h_k^{(m,n)}\!|\!e^{j(\!2\pi\frac{l\!-\!k}{T_s}t + \text{arg}(\! h_l^{(m,n)} \!) - \text{arg}(\! h_k^{(m,n)} \!)\!)} \nonumber\\
        &+ \!\!\! \sum_{k=1}^{N_f} \!\! 2\!A_r|\!h_k^{(m,n)}\!| \! \cos{\!\!\left(\!\! 2\pi(\!f_c\!\!-\!\!f_r\!\!+\!\!\frac{k\!\!-\!\!1}{T_s}\!)t \!+\! \text{arg}(h_k^{(m,n)}\!) \!\!\!\right)}\!. \label{eqIsum}
\end{align}
Eq. (\ref{eqIsum}) reveals that the hologram is a complex superposition of weighted sinusoidal signals in the wideband scenario. Although the amplitude of the superimposed signal is the only information we have acquired, the hologram contains the phase difference between the object wave and the reference wave. It indicates that the reference wave serves as a baseline, enabling the recovery of the object wave and estimation of CSI with multiple holograms.

The nonlinearity of holograms manifests in two primary aspects. First, this characteristic renders most powerful and well-established linear signal analysis tools ineffective for the holographic channel estimation problem. It is worth noting that the nonlinear property of holograms leads to signal overlapping in the frequency domain. This property makes it quite challenging to directly apply classical signal analysis tools to hologram sequences, such as the matrix pencil (MP) \cite{Hua:92TSP}, estimation of signal parameters via rotational invariance techniques (ESPRIT)\cite{Roy:86TASSP}, and the discrete Fourier transform (DFT), which are essentially built on linear structures. Second, the nonlinearity of holograms is a unique type that can be mitigated or transformed, thereby offering a trade-off between complexity and accuracy. It is possible to construct a special transformation function or mapping function to recover holograms back into a linear sequence, which is discussed below. Such a transformation process enables the use of the above classical tools for holographic channel estimation, thereby making it possible to achieve an effective trade-off between time complexity and estimation accuracy.

\subsection{Holographic Channel Recovery} \label{secHoloComm_HCR}
The hologram is a nonlinear function of the phase and amplitude of the object wave. Moreover, the number of summation terms in the hologram grows quadratically as the number of subcarriers increases. Therefore, it is relatively complicated to estimate the object wave by directly solving the system of nonlinear equations constructed from Eq. (\ref{eqIsum}). Accordingly, we extend the idea of phase shifting interferometry (PSI) \cite{Meng:06OL}, \cite{Ichirou:97OL} to recover the object wave based on multiple holograms.
 
 Since the system uses OFDM frequency set, it is obvious that $E_o(t+T_s) = e^{j2\pi f_c T_s}E_o(t)$. When the time interval between two adjacent samples is set to $T_s$, the hologram satisfies
 \begin{align}
    E_I(t+T_s) 
        &= |E_r(t+T_s) + E_o(t+T_s)|^2 \nonumber\\
        &= |e^{j2\pi f_r T_s}E_r(t) + e^{j2\pi f_c T_s}E_o(t)|^2 \nonumber\\              
        &= |e^{j\delta}E_r(t) + E_o(t)|^2, \label{eqI2}
 \end{align}
 where $\delta = 2\pi(f_r - f_c)T_s$ is the phase difference between the reference wave and the object wave. 
 
 Comparing Eq. (\ref{eqI1}) and Eq. (\ref{eqI2}), we may observe that although a single hologram only provides intensity information, the pre-designed hologram sequence contains the phase difference. Therefore, we can recover the complex domain representation of $E_o(t)$ from the intensity domain utilizing at least two holograms, which is discussed in detail below.

\begin{figure*}[!t]
    \centering
    \subfloat[]{\includegraphics[width=\columnwidth]{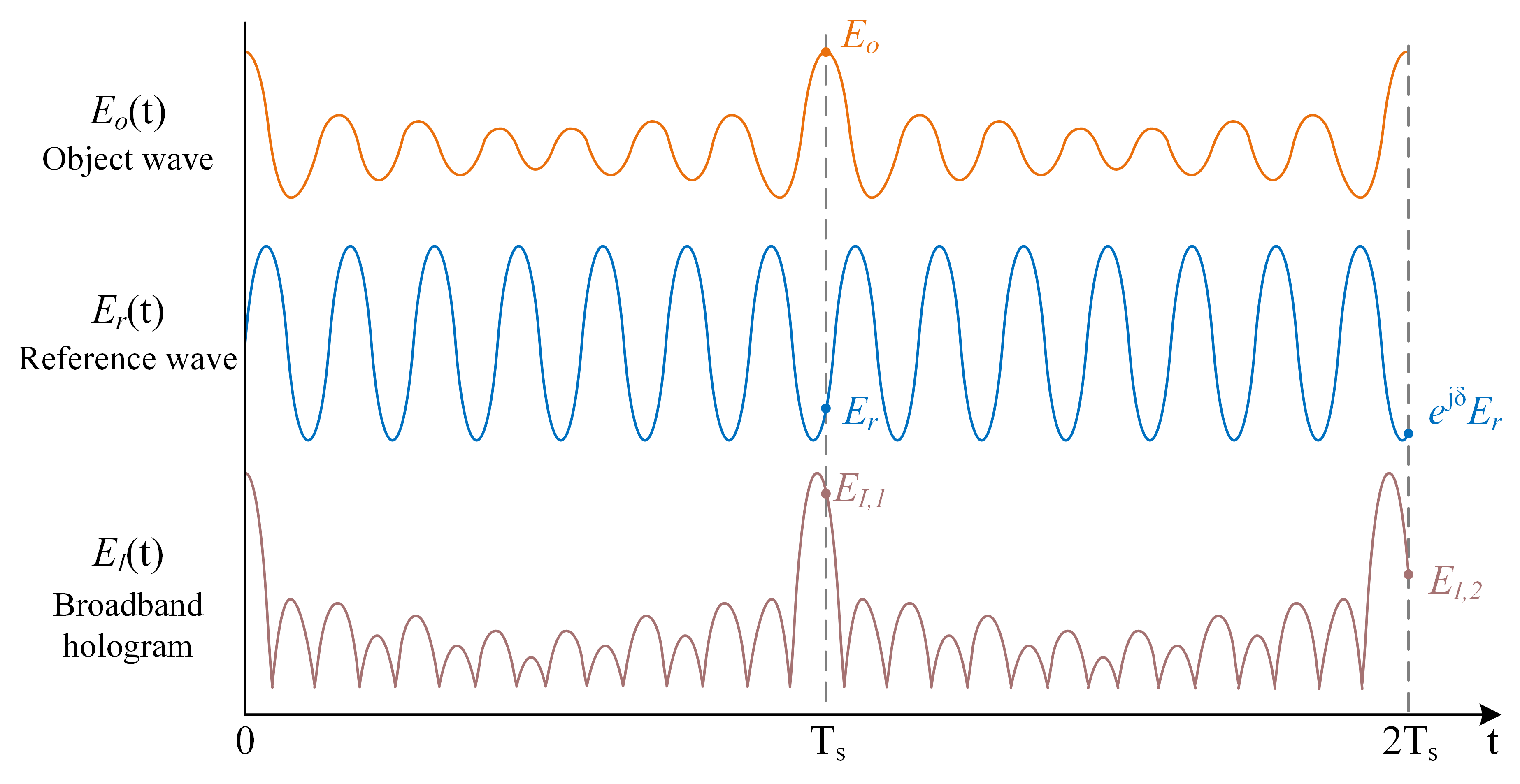} \label{figHCR_t}}
    \hfil
    \subfloat[]{\includegraphics[width=\columnwidth]{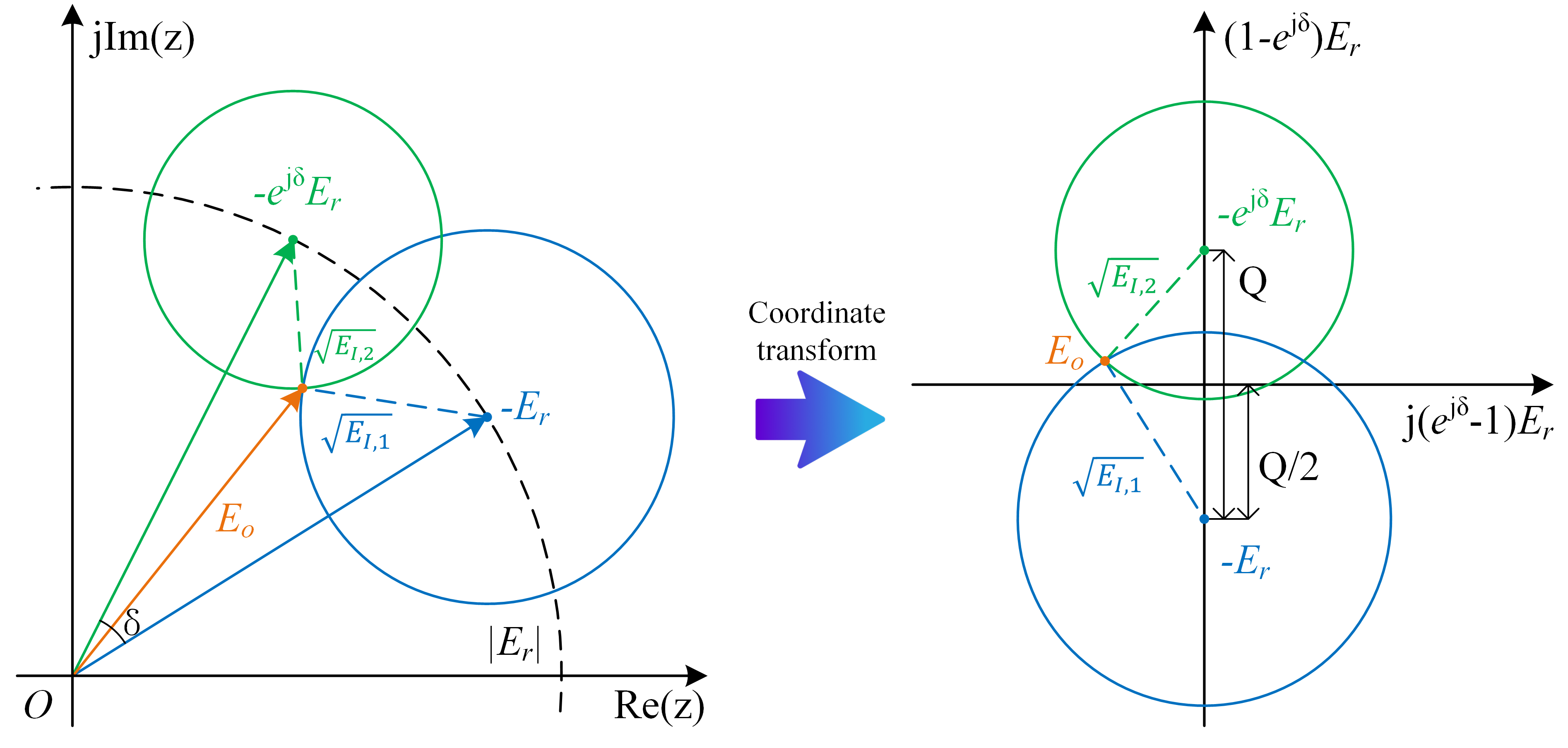} \label{figHCR_c}}
    \caption{Representation of holographic channel recovery in different domains. (a) Time domain. (b) Complex domain.}
    \label{figHCR}
\end{figure*}

For ease of notation, we denote $E_I(t)$ by $E_{I,1}$ and $E_I(t+T_s)$ by $E_{I,2}$. Since the hologram is a function of signal power and phase difference between $E_r(t)$ and $E_o(t)$, Eq. (\ref{eqI1}) and Eq. (\ref{eqI2}) have the following equivalent representation
\begin{align}
    E_{I,1} 
        &= A_o^2 + A_r^2 + 2A_oA_r\cos{\phi},\\
    E_{I,2} 
        &= A_o^2 + A_r^2 + 2A_oA_r\cos{(\phi - \delta)} \nonumber\\
        &= A_o^2 + A_r^2 + 2A_oA_r\left[ \cos{\phi}\cos{\delta} + \sin{\phi}\sin{\delta} \right],
\end{align}
where $A_o = |E_o(t)|$ and $\phi = \text{arg}(E_o(t)) - 2\pi f_r t$. From the above equations, we may observe
\begin{align}
    A_o\cos{\phi} &= \frac{E_{I,1} - (A_o^2 + A_r^2)}{2A_r}, \label{eqAcos}\\
    A_o\sin{\phi} &= \frac{E_{I,2} - E_{I,1}\cos{\delta} - (1-\cos{\delta})(A_o^2 + A_r^2)}{2A_r\sin{\delta}}. \label{eqAsin}
\end{align}
Using the relation $(A_o\cos{\phi})^2 + (A_o\sin{\phi})^2 = A_o^2$, we can find that Eq. (\ref{eqAcos}) and Eq. (\ref{eqAsin}) form the following quadratic equation with respect to $b = A_o^2 + A_r^2$,
\begin{align}
    u'b^2 + v'b + w' = 0,
\end{align}
where
\begin{align}
    u' &= 2-2\cos{\delta},\\
    v' &= -2(1-\cos{\delta})(E_{I,1} + E_{I,2}) - 4A_r^2\sin^2{\delta},\\
    w' &= E_{I,1}^2 + E_{I,2}^2 - 2E_{I,1}E_{I,2}\cos{\delta} + 4A_r^4\sin^2{\delta}.
\end{align}
It is convenient to verify that $v'^2 - 4u'w' \geq 0$ and the equation holds a solution. Considering $A_r > A_o$, the estimation of the object wave is given by
\begin{align}
    \hat{E}_o(t) \!
        &= \! A_o e^{j(\phi + 2\pi f_r t)} \nonumber \\
        &= \! (A_o\cos{\phi} + jA_o\sin{\phi})e^{j2\pi f_r t} \nonumber \\
        &= \!\! \left[ \! \frac{E_{I\!,1} \!\! - \! b}{2A_r} \!\! + \! j\frac{E_{I\!,2} \! - \! E_{I\!,1}\!\cos{\delta} \! - \! (\!1 \!\! - \!\! \cos{\delta}\!) b}{2A_r\sin{\delta}} \! \right] \!\! e^{j2\pi f_r t}\!, \label{eqPSI}
\end{align}
where 
\begin{align}
    b = \frac{-v'-\sqrt{v'^2-4u'w'}}{2u'}. \label{eqb}
\end{align}

Note that $E_{I,1}$ and $E_{I,2}$ denote the Euclidean distance from $E_o(t)$ to $-E_r(t)$ and $E_o(t)$ to $-e^{j\delta} E_r(t)$ respectively. Thus, in the complex plane, the object wave lies at the intersection of two circles. Assuming $|E_r(t)| > |E_o(t)|$, the object wave is supposed to be located at the intersection closer to the origin, as is shown in Fig. \ref{figHCR}\subref{figHCR_c}. Considering the physical significance of Eq. (\ref{eqI1}) and Eq. (\ref{eqI2}), the problem of estimating object wave is equivalent to finding the intersection of two circles, which is given by
\begin{align}
    \left \{ 
    \begin{aligned}
        &(\mathfrak{R}(E_o) \!+\! \mathfrak{R}(E_r))^2 \!+\! (\mathfrak{I}(E_o) \!+\! \mathfrak{I}(E_r))^2 \!=\! E_{I,1} \\
        &(\mathfrak{R}(E_o) \!+\! \mathfrak{R}(e^{j\delta} E_r))^2 \!+\! (\mathfrak{I}(E_o) \!+\! \mathfrak{I}(e^{j\delta} E_r))^2 \!=\! E_{I,2}
    \end{aligned} \right. . \label{eqGeo}
\end{align}
The above system of quadratic equations is solved by transforming the coordinate system, which leads to another equivalent expression for the estimation of object wave. The estimation formula equivalent to Eq. (\ref{eqPSI}) is expressed as
\begin{align}
    \hat{E}_o(t) = \left[ (1 - e^{j\delta})(s_1(t) - j s_2(t)) - \frac{e^{j\delta} + 1}{2} \right] E_r(t), \label{eqPSIG}
\end{align}
where
\begin{align}
    s_1(t) &= \frac{E_{I,1} - E_{I,2}}{2 Q^2}, \\
    s_2(t) &= \left( \frac{E_{I,1} + E_{I,2}}{2 Q^2} - \frac{(E_{I,1} - E_{I,2})^2}{4 Q^4} - \frac{1}{4} \right)^{\frac{1}{2}}, \\
    Q   &= 2 \left| \sin{(\delta/2)} E_r(t) \right| = 2 \left| \sin{(\delta/2)} A_r \right|.
\end{align}

Fig. \ref{figHCR} \subref{figHCR_c} illustrates the physical meaning of holographic channel recovery, i.e., Eq. (\ref{eqPSIG}), in the complex plane. We have a series of observation points, along with the distance between each point and the target. Holography essentially predicts the movement trend of the target on a two-dimensional plane based on feedback from these observation points, which is highly similar to the localization problem. For coherent signals, the problem is relatively straightforward and has been fully addressed in optical holography. In contrast, for incoherent signals, the resulting holograms are time-varying, necessitating the design of a sampling strategy.

Although Eq. (\ref{eqPSI}) and Eq. (\ref{eqPSIG}) are identical in mathematics, Eq. (\ref{eqPSIG}) provides a more intuitive reflection of the relationship among the object wave, the reference wave and the hologram. Therefore, we primarily consider Eq. (\ref{eqPSIG}) as the object wave recovery formula in the following. It can be seen from Eq. (\ref{eqPSI}) and Eq. (\ref{eqPSIG}) that the complex domain representation of the received signal is estimated from the intensity domain utilizing only two holograms within two OFDM symbol periods. In our architecture, combiners and power detection devices replace complex operations in traditional systems such as ADC, filtering and down conversion, offering a new vision for the efficient deployment of ultra-large-scale arrays.

\section{Holographic Channel Estimation} \label{secHCE}
In this section, two hologram-based channel estimation algorithms will be proposed in Sec. \ref{secHCE_MLE} and \ref{secHCE_GROWS} respectively. Then, we derive the CRLB for the holographic channel estimation problem in Sec. \ref{secHCE_CRLB}. Note that in this section, we consider the $(m,n)$-th unit of the HIS and drop the superscript $(m,n)$ for ease of notation.

\subsection{Maximum Likelihood Estimation} \label{secHCE_MLE}
Assume the electromagnetic noise $\omega(t) \sim \mathcal{CN}(0,\sigma_{\omega}^2)$. Since the hologram is the power of the superimposed signal, the distribution of the noisy hologram $\tilde{E}_I(t) = |E_r(t) + E_o(t) + \omega(t)|^2$ is characterized by a non-central chi-squared distribution $\mathcal{X}_2^2(|\mu(t)|^2, \sigma_{\omega}^2/2)$. The mean value $\mu(t)$ is given by
\begin{align}
    \mu(t) = E_r(t) + E_o(t)
           = E_r(t) + \mathbf{\Phi}(t)^T \mathbf{h},
\end{align}
where $\mathbf{\Phi}(t) \!=\! [e^{j2\pi f_1t}, \!\cdots, \!e^{j2\pi f_{N_f}t}]^T$ and $\mathbf{h} \!=\! [h_1, \!\cdots,\! h_{N_f}]^T$. Thus, the probability density function of $\tilde{E}_I(t)$ is given by
\begin{align}
    f(x) \!\!=\!\! \frac{1}{\sigma_{\omega}^2} \! \exp\left(-\frac{x + |\mu(t)|^2}{\sigma_{\omega}^2}\right) \!\! I_0\left(\frac{2\sqrt{x}|\mu(t)|}{\sigma_{\omega}^2}\right)\!\!, x \geq 0,
\end{align}
where $I_0(\cdot)$ is the zeroth-order modified Bessel function of the first kind. Therefore, the log-likelihood function of $\mathbf{h}$
 based on the holograms $\tilde{E}_I[l], l = 1, \cdots, L$ is expressed as
 \begin{align}
    \mathcal{F}(\mathbf{h}) \!\!
        = \!\!\! \sum_{l=1}^L \!\! \left[ \! \log\!{I_0 \!\!\left(\!\!\! \frac{2 \! \sqrt{\tilde{E}_I[l]} |\mu_l|}{\sigma_{\omega}^2} \!\!\!\right)} \!\!-\! \frac{\tilde{E}_I[l] \!+\! |\mu_l|^2}{\sigma_{\omega}^2} \!\right] \!\!\!-\!\! L \! \log{ \! \sigma_{\omega}^2},
 \end{align}
where $\mu_l = \mu(t_l)$. Under the maximum likelihood criterion, the value of $\mathbf{h}$ is required to maximize the log-likelihood function $\mathcal{F}(\mathbf{h})$. This unconstrained optimization problem can be solved by complex Newton's method, which requires the gradient and Hessian matrix of $\mathcal{F}(\mathbf{h})$. Hence, the partial derivative of $\mathcal{F}(\mathbf{h})$ is derived below.

According to the definition of Wirtinger partial derivative $\frac{\partial}{\partial z^*} = \frac{1}{2} \left( \frac{\partial}{\partial \mathfrak{R}(z)} + j\frac{\partial}{\partial \mathfrak{I}(z)} \right)$, we may obtain
\begin{align}
    \frac{\partial |\mu_l|^2}{\partial h_m^*} 
		= \frac{1}{2} \left( \frac{\partial |\mu_l|^2}{\partial \mathfrak{R}(h_m)} + j\frac{\partial |\mu_l|^2}{\partial \mathfrak{I}(h_m)} \right) 
		= \mu_l \Phi_{l,m}^*,
\end{align}
where $\Phi_{l,m} = e^{j2\pi f_m t_l}$. Similarly, the derivative of $|\mu_l|$ is given by
\begin{align}
    \frac{\partial |\mu_l|}{\partial h_m^*}
		=& \frac{1}{2} \left( \frac{\partial |\mu_l|}{\partial \mathfrak{R}(h_m)} + j\frac{\partial |\mu_l|}{\partial \mathfrak{I}(h_m)} \right) \nonumber \\
		=& \frac{1}{2} \!\! \left( \! \frac{1}{2 |\mu_l|} \frac{\partial |\mu_l|^2}{\partial \mathfrak{R}(h_m)} \!+\! j \frac{1}{2 |\mu_l|} \frac{\partial |\mu_l|^2}{\partial \mathfrak{I}(h_m)} \! \right) \!
		= \! \frac{\mu_l \Phi_{l,m}^*}{2 |\mu_l|}.
\end{align}
Given $z_l = 2\sqrt{\tilde{E}_I[l]}|\mu_l|/\sigma_{\omega}^2$, we may obtain the derivative of $\log{I_0(z_l)}$ based on the chain rule for partial derivatives in complex functions as
\begin{align}
    \frac{\partial \! \log{\! I_0(\!z_l\!)}}{\partial h_m^*} \!
		=& 2 \frac{\partial \log{I_0(z_l)}}{\partial z_l} \frac{\partial z_l}{\partial h_m^*} \nonumber \\
		=& \frac{4 I_1(\!z_l\!) \! \sqrt{\tilde{E}_I[l]}}{I_0(\!z_l\!) \sigma_{\omega}^2} \frac{\partial |\mu_l|}{\partial h_m^*}
		\!=\!\! \frac{2 \! \sqrt{\!E_I\![l]} R(\!z_l\!)}{\sigma_{\omega}^2 |\mu_l|}\mu_l \Phi_{l,m}^*,
\end{align}
where $I_1(\cdot)$ is the first-order modified Bessel function of the first kind and $R(z_l) = I_1(z_l)/I_0(z_l)$. From the above calculations, it follows that 
\begin{align}
    \frac{\partial \mathcal{F}(\mathbf{h})}{\partial h_m^*}
		= \frac{1}{\sigma_{\omega}^2} \sum_{l=1}^L \left( \frac{2\sqrt{\tilde{E}_I[l]}}{|\mu_l|} R(z_l) - 1 \right) \mu_l \Phi_{l,m}^*. \label{eqGradm}
\end{align}
Accordingly, the complex gradient of $\mathcal{F}(\mathbf{h})$ is given by
\begin{align}
    \nabla_{\mathbf{h}^*} \mathcal{F}(\mathbf{h})
		= \frac{\partial \mathcal{F}(\mathbf{h})}{\partial \mathbf{h}^*} 
		= \mathbf{\Phi}^* \mathbf{R}_1(\mathbf{h}) \boldsymbol{\mu},
\end{align}
where
\begin{align}
    \boldsymbol{\mu} 
        =& [\mu_1, \cdots,\mu_L]^T  \in \mathbb{C}^{L} , \\
    \mathbf{\Phi} 
        =& [\mathbf{\Phi}(t_1), \cdots,\mathbf{\Phi}(t_L)]  \in \mathbb{C}^{N_f \times L} , \\
    \mathbf{R}_1(\mathbf{h}) 
        =& \text{diag} \left\{ \frac{z_l R(z_l)}{|\mu_l|^2} - \frac{1}{\sigma_{\omega}^2} \right\}  \in \mathbb{C}^{L \times L}.
\end{align}

Based on the definition of the complex Hessian matrix, the partial Hessian matrices of $\mathcal{F}(\mathbf{h})$ are given by
\begin{align}
	\boldsymbol{\mathcal{H}}_{\mathbf{h}, \mathbf{h}} (\mathcal{F}(\mathbf{h}))
		=& \frac{\partial^2 \mathcal{F}(\mathbf{h})}{\partial \mathbf{h} \partial \mathbf{h}^T}
		= \mathbf{\Phi} \mathbf{R}_3(\mathbf{h}) \mathbf{\Phi}^T, \\
	\boldsymbol{\mathcal{H}}_{\mathbf{h}, \mathbf{h}^*} (\mathcal{F}(\mathbf{h}))
		=& \frac{\partial^2 \mathcal{F}(\mathbf{h})}{\partial \mathbf{h} \partial \mathbf{h}^H}
		= \mathbf{\Phi} \mathbf{R}_2(\mathbf{h}) \mathbf{\Phi}^H, \\
    \boldsymbol{\mathcal{H}}_{\mathbf{h}^*, \mathbf{h}} (\mathcal{F}(\mathbf{h}))
		=& \frac{\partial^2 \mathcal{F}(\mathbf{h})}{\partial \mathbf{h}^* \partial \mathbf{h}^T}
		= \mathbf{\Phi}^* \mathbf{R}_2(\mathbf{h}) \mathbf{\Phi}^T, \\
    \boldsymbol{\mathcal{H}}_{\mathbf{h}^*, \mathbf{h}^*} (\mathcal{F}(\mathbf{h}))
		=& \frac{\partial^2 \mathcal{F}(\mathbf{h})}{\partial \mathbf{h}^* \partial \mathbf{h}^H}
		= \mathbf{\Phi}^* \mathbf{R}_3(\mathbf{h})^* \mathbf{\Phi}^H,
\end{align}
where 
\begin{align}
    \mathbf{R}_2(\mathbf{h}) 
        &= \frac{1}{\sigma_{\omega}^4} \text{diag} \{ 2 \tilde{E}_I[l] - 2 \tilde{E}_I[l] R^2(z_l) - \sigma_{\omega}^2 \}, \\
    \mathbf{R}_3(\mathbf{h}) 
        &= \text{diag} \left\{\frac{ z_l^2 - z_l^2 R^2(z_l) - 2 z_l R(z_l)}{2 \mu_l^2} \right\}.
\end{align}
The detailed derivation of the Hessian matrix can be found in Appendix \ref{apxHess}. 

Utilizing the gradient and Hessian matrix of $\mathcal{F}(\mathbf{h})$, the step size in the Newton iteration is given by
\begin{align}
    \begin{bmatrix}
        \boldsymbol{\mathcal{H}}_{\mathbf{h}^*, \mathbf{h}} (\!\mathcal{F}(\!\mathbf{h}_k\!)\!) \!\!&\!\!\! \boldsymbol{\mathcal{H}}_{\mathbf{h}^*, \mathbf{h}^*} (\!\mathcal{F}(\!\mathbf{h}_k\!)\!) \\
		\boldsymbol{\mathcal{H}}_{\mathbf{h}, \mathbf{h}} (\!\mathcal{F}(\!\mathbf{h}_k\!)\!)   \!\!&\!\!\! \boldsymbol{\mathcal{H}}_{\mathbf{h}, \mathbf{h}^*} (\!\mathcal{F}(\!\mathbf{h}_k\!)\!)
	\end{bmatrix} \!\!\!
	\begin{bmatrix}
		\Delta \! \mathbf{h}_{k} \\ 
		\Delta \! \mathbf{h}_{k}^*
	\end{bmatrix} \!\!=\!\! - \!\!
	\begin{bmatrix}
		\nabla_{\mathbf{h}} \mathcal{F}(\!\mathbf{h}_k\!) \\
		\nabla_{\mathbf{h}^*} \mathcal{F}(\!\mathbf{h}_k\!)
	\end{bmatrix}\!\!, \label{eqDeltah}
\end{align}
where $\nabla_{\mathbf{h}} \mathcal{F}(\mathbf{h}_k) = (\nabla_{\mathbf{h}^*} \mathcal{F}(\mathbf{h}_k))^*$. Then the updating formula of Newton's method \cite{Casella:21CL} is given by
\begin{align}
	\begin{bmatrix}
		\mathbf{h}_{k+1} \\
		\mathbf{h}_{k+1}^*
	\end{bmatrix} = 
	\begin{bmatrix}
		\mathbf{h}_{k} \\
		\mathbf{h}_{k}^*
	\end{bmatrix} + q
	\begin{bmatrix}
		\Delta \mathbf{h}_{k} \\ 
		\Delta \mathbf{h}_{k}^*	
	\end{bmatrix},
\end{align}
where $q$ is the step size of each iteration. In order to ensure a sufficient increase in the log-likelihood function during the iteration, it is necessary for the step size to satisfy the following Armijo condition \cite{Boyd:04CO}:
\begin{align}
    \mathcal{F}\!(\!\mathbf{h}_{k} \!\!+\!\! q \! \Delta \! \mathbf{h}_{k} \!)
        \!\!<\!\! \mathcal{F}\!(\mathbf{h}_{k}\!)\!\! + \!\! \alpha q \! \left(\! \nabla_{\!\mathbf{h}} \mathcal{F}\!(\mathbf{h}_{k}\!)\!^T \!\! \Delta \! \mathbf{h}_{k} \!\!+\!\! \nabla_{\!\mathbf{h}^*} \! \mathcal{F}\!(\mathbf{h}_{k}\!)\!^T \!\!\Delta \! \mathbf{h}_{k}^*\!\right)\!\!. \label{eqArmijo}
\end{align}
If the Armijo condition is not satisfied, $q$ needs to be adjusted by a reduction factor $\beta \in (0,1)$ as $q \gets \beta q$. 

\begin{algorithm}[H]
\setstretch{1.35}
\caption{Wideband Hologram-based ML Estimation}
\label{algMLE}
\begin{algorithmic}[1]
\REQUIRE Armijo factor $\alpha \in (0, 0.5)$, reduction factor $\beta \in (0,1)$, holograms $\tilde{E}_I[l]$ and noise variance $\sigma_{\omega}^2$.
\ENSURE $\hat{\mathbf{h}}$.
\STATE Obtain an initial estimation $\hat{\mathbf{h}}$. 
\WHILE{$\hat{\mathbf{h}}$ not convergent}
\STATE Calculate $\Delta \hat{\mathbf{h}}$ based on Eq. (\ref{eqDeltah}).
\STATE $q \gets 1$.
\WHILE{Eq. (\ref{eqArmijo}) is not satisfied}
\STATE $q \gets \beta q$.
\ENDWHILE
\STATE $\hat{\mathbf{h}} \gets \hat{\mathbf{h}} + q \Delta \hat{\mathbf{h}}$.
\ENDWHILE
\RETURN $\hat{\mathbf{h}}$
\end{algorithmic}
\end{algorithm}

The whole WH-ML estimation is summarized in Algorithm \ref{algMLE}. In this hologram-based ML estimation problem, the channel state matrix $\mathbf{\hat{h}}$ is optimized by the complex Newton's method. Moreover, the Armijo condition is considered to adjust the step size $\Delta{\hat{\mathbf{h}}}$ of each iteration for the sufficient increase of $\mathcal{F}(\hat{\mathbf{h}})$.
The storage space requirements of WH-ML primarily originate from the Hessian matrix computation and the Newton step size calculation in each iteration. Therefore, based on the space complexity of matrix multiplication and linear system solving, the space complexity of WH-ML can be expressed as $\mathcal{O}(N_t L) + \mathcal{O}(N_t N_f L) + \mathcal{O}(N_t N_f^2)$.

The WH-ML method treats the hologram as a generalized nonlinear function for directly estimating the CSI. Consequently, WH-ML essentially represents a direct solution to a system of nonlinear equations, employing the complex Newton's method. In Sec. \ref{secHCE_GROWS}, we introduce another holographic channel estimation algorithm based on the idea of nonlinearity cancellation. This algorithm constructs a mapping function from nonlinear sequences to linear sequences, thereby achieving a well-balanced trade-off between time complexity and estimation accuracy.

\subsection{Geometric Rotation-based Object Wave Sensing} \label{secHCE_GROWS}
Although the ML estimator can achieve the CRLB asymptotically \cite{Myung:03JMP}, it may be trapped in locally optimal solutions with complicated computation due to the optimization algorithm employed and the complexity of the wideband holographic channel estimation problem. Therefore, a generalized wideband holographic channel estimation algorithm with much lower complexity is explored in the following. In Sec. \ref{secHoloComm_HCR}, we have analyzed the method of recovering the object wave from holograms. Assume that $L$ equally-spaced observations $\tilde{E}_I[l] = E_I(t_l)$ with instants $t_l = \frac{l}{L}T_s, l = 0, \cdots, L-1$ are sampled in a symbol period. Considering that Eq. (\ref{eqPSI}) utilizes two holograms separated by a time interval of $T_s$, $2L$ hologram samples are required to recover the object wave at $L$ time slots. Therefore, we utilize hologram samples within two symbol periods in the following. Utilizing $2L$ holograms, the sample sequence of the object wave is estimated as
\begin{align}
    \hat{E}_o[l] 
        \!=\! \hat{E}_o(t_l)
        \!=\!\! \left[\! (1 \!-\! e^{j\delta})\!(s_1[l] \!-\! j s_2[l]) \!-\! \frac{e^{j\delta} \!+\! 1}{2} \!\right] \!\! E_r[l], \label{eqPSIl}
\end{align}
where $s_1[l] = s_1(t_l)$, $s_2[l] = s_2(t_l)$ and $E_r[l] = E_r(t_l)$. Apply $L$-point DFT to the sequence $p[l] = \hat{E}_o[l] e^{-j2\pi \frac{f_c T_s-1}{L}l}$ yields
\begin{align}
    P(k) = \sum_{l=0}^{L-1} p[l]e^{-j2\pi \frac{k}{L}l} = L h_k.
\end{align}
Then the frequency component $h_k$ is estimated as
\begin{align}
    \hat{h}_k = \frac{P(k)}{L} = \frac{\sum_{l=0}^{L-1} p[l]e^{-j2\pi \frac{k}{L}l}}{L}.
\end{align}

The whole geometric rotation-based object wave sensing algorithm is summarized in Algorithm \ref{algGROWS}. The complexity of this algorithm is analyzed below. Note that the estimation of $E_o[l]$ in Algorithm \ref{algGROWS} is independent of other iterations and units. Therefore, the calculations can be performed in parallel for each unit, resulting in an $\mathcal{O}(1)$ time complexity. The complexity of this algorithm is dominated by the computation of fast Fourier transform (FFT). It can be verified that the wideband channel recovery has a complexity order of $\mathcal{O}(L) + \mathcal{O}(L \log{(L)})$. For an array with $N_t$ antenna units, the calculation complexity of GROWS algorithm is $\mathcal{O}(L) + \mathcal{O}(L N_t \log{(N_t)}) + \mathcal{O}(L N_t \log{(L N_t)})$.
It should be noted that the channel recovery process involves only element-wise operations between hologram sequences and supports parallel processing across units. Taking into account the in-place computation property of FFT, the spatial complexity of the GROWS algorithm is $\mathcal{O}(L N_t)$.

\begin{algorithm}[H]
\setstretch{1.35}
\caption{Geometric Rotation-based Object Wave Sensing}
\label{algGROWS}
\begin{algorithmic}[1]
\REQUIRE Symbol period $T_s$, carrier frequency $f_c$, reference wave $E_r[l]$ and holograms $\tilde{E}_I[l]$.
\ENSURE $\hat{\mathbf{h}}$.
\STATE $\delta \gets 2\pi (f_r - f_c) T_s$, $Q \gets 2 |\sin{(\delta/2)} A_r|$.
\FOR{$l = 0, \cdots, L-1$}
\STATE $s_1[l] \gets \frac{\tilde{E}_I[l] - \tilde{E}_I[l + L]}{2 Q^2}$.
\STATE $s_2[l] \gets (\frac{\tilde{E}_I[l] + \tilde{E}_I[l + L]}{2 Q^2} - \frac{(\tilde{E}_I[l] - \tilde{E}_I[l + L])^2}{4 Q^4} - \frac{1}{4})^{\frac{1}{2}}$.
\STATE $\hat{E}_o[l] \gets \left[ (1 - e^{j\delta})(s_1[l] - j s_2[l]) - \frac{e^{j\delta} + 1}{2} \right] E_r[l]$.
\STATE $p[l] \gets \hat{E}_o[l] e^{-j2\pi \frac{f_c T_s-1}{L}l}$.
\ENDFOR
\STATE $\hat{\mathbf{h}} \gets \text{FFT}(p)/L$.
\RETURN $\hat{\mathbf{h}}$
\end{algorithmic}
\end{algorithm}

\subsection{Cram\'er-Rao Lower Bound}\label{secHCE_CRLB}
To measure the attainable precision of the wideband holographic channel estimation, we derive the complex CRLB \cite{Ollila:08} in \textbf{Theorem \ref{thmCRLB}}.

\begin{thm} \label{thmCRLB}
    For the problem of channel estimation with $L$ holograms under the non-central chi-squared distribution, the CRLB is given by
    \begin{align}
        \text{CRLB}(\mathbf{h})
            = \begin{bmatrix}
                \boldsymbol{\mathcal{I}}_{\mathbf{h}}   & \boldsymbol{\mathcal{P}}_{\mathbf{h}} \\
			    \boldsymbol{\mathcal{P}}_{\mathbf{h}}^* & \boldsymbol{\mathcal{I}}_{\mathbf{h}}^* 
            \end{bmatrix}^{-1}. 
    \end{align}
    The complex information matrix and the pseudo-information matrix are expressed as 
    \begin{align}
        \boldsymbol{\mathcal{I}}_{\mathbf{h}} 
		  =& \frac{1}{\sigma_{\omega}^4} \mathbf{\Phi}^* \left( \boldsymbol{\mu} \boldsymbol{\mu}^H + \boldsymbol{\mathcal{J}}_{\mathcal{I}} \right) \mathbf{\Phi}^T, \\
        \boldsymbol{\mathcal{P}}_{\mathbf{h}} 
		  =& \frac{1}{\sigma_{\omega}^4} \mathbf{\Phi}^* \left( \boldsymbol{\mu} \boldsymbol{\mu}^T + \boldsymbol{\mathcal{J}}_{\mathcal{P}} \right) \mathbf{\Phi}^H,
    \end{align}
    where
    \begin{align}
        \boldsymbol{\mathcal{J}}_{\mathcal{I}} 
            =& \text{diag}\{ 4 \left( J(\gamma_l)-1 \right) |\mu_l|^2 \}, \\
        \boldsymbol{\mathcal{J}}_{\mathcal{P}} 
            =& \text{diag}\{ 4 \left( J(\gamma_l)-1 \right) \mu_l^2 \}.
    \end{align}
    The function $J(\gamma_l)$ is defined as
    \begin{align}
        J(\gamma_l)
            \!\!=\!\!\! \int_{0}^{+\infty} \!\!\!\!\!\!\! \gamma_l t \exp\!{(-\gamma_l (1+t)\!)} I_0 \!\left( 2 \gamma_l \sqrt{t} \right) \! R^2 \!\!\left(\! 2 \gamma_l \sqrt{t} \! \right)\! dt. \label{eqJgamma}
    \end{align}
\end{thm}
\begin{IEEEproof}
    The proof can be found in Appendix \ref{apxThmCRLB}.
\end{IEEEproof}
\begin{rem}
    Defining $\boldsymbol{\mathcal{R}}_{\mathbf{h}} = \boldsymbol{\mathcal{I}}_{\mathbf{h}} - \boldsymbol{\mathcal{P}}_{\mathbf{h}} (\boldsymbol{\mathcal{I}}_{\mathbf{h}}^{-1})^* \boldsymbol{\mathcal{P}}_{\mathbf{h}}^*$, it follows from Corollary 2 in \cite{Ollila:08} that the matrix $\text{Cov}(\mathbf{t}) - \boldsymbol{\mathcal{R}}_{\mathbf{h}}^{-1}$ is positive semidefinite, if $\mathbf{t}$ is an unbiased estimator of $\mathbf{h}$. Since the calculation of $J(\gamma)$ is relatively complicated, there is an approximate expression for $J(\gamma)$ according to \cite{Zhu:23TIT}. Due to the asymptotic equation $x R^2(2\sqrt{x}) \approx x - \frac{\sqrt{x}}{2}$, it follows that
    \begin{align}
        \frac{\tilde{E}_I[l]}{|\mu_l|^2} R^2\left( \frac{2\sqrt{\tilde{E}_I[l]}|\mu_l|}{\sigma_{\omega}^2} \right)
            \approx \frac{\tilde{E}_I[l]}{|\mu_l|^2} - \frac{\sigma_{\omega}^2 \sqrt{\tilde{E}_I[l]}}{2 |\mu_l|^3}.
    \end{align}
    Therefore, $J(\gamma_l)$ is approximated by
    \begin{align}
        J(\gamma_l) 
            &= \mathbb{E}\left[ \frac{\tilde{E}_I[l]}{|\mu_l|^2} R^2\left( \frac{2\sqrt{\tilde{E}_I[l]}|\mu_l|}{\sigma_{\omega}^2} \right) \right] \nonumber \\
            &\approx \frac{1}{|\mu_l|^2} \mathbb{E}[\tilde{E}_I[l]] - \frac{\sigma_{\omega}^2}{2|\mu_l|^3} \mathbb{E}\left[ \sqrt{\tilde{E}_I[l]} \right]. \label{eqJapprox}
    \end{align}
    Since $\tilde{E}_I[l]$ follows the non-central chi-squared distribution, $\sqrt{\tilde{E}_I[l]}$ follows the non-central chi distribution, i.e., $\tilde{E}_I[l] \sim \sqrt{\sigma_{\omega}^2/2} \mathcal{X}_2 \left( \sqrt{2} |\mu_l|/\sigma_{\omega} \right)$. Accordingly, the expectation of $\tilde{E}_I[l]$ and $\sqrt{\tilde{E}_I[l]}$ are given by
    \begin{align}
        \mathbb{E}[\tilde{E}_I[l]] \!\!
            =& \sigma_{\omega}^2 + |\mu_l|^2, \label{eqEChiSq} \\
        \mathbb{E}[\sqrt{\tilde{E}_I[l]}] \!\!
            =& \frac{\sqrt{\pi} \sigma_{\omega}}{2} \exp{\left( -\frac{|\mu_l|^2}{2 \sigma_{\omega}^2}  \right)} \nonumber \\
             &\times \!\! \left[ \!\! \left( \! 1 \!\!+\!\! \frac{|\mu_l|^2}{\sigma_{\omega}^2} \! \right) \! I_0 \!\! \left( \! \frac{|\mu_l|^2}{2 \sigma_{\omega}^2} \! \right) \!+\! \frac{|\mu_l|^2}{\sigma_{\omega}^2} \! I_1 \!\! \left( \! \frac{|\mu_l|^2}{2 \sigma_{\omega}^2} \! \right) \!\! \right] \!. \label{eqEChi}
    \end{align}
    Substituting Eq. (\ref{eqEChiSq}) and Eq. (\ref{eqEChi}) into Eq. (\ref{eqJapprox}), the approximate expression $\hat{J}(\gamma_l)$ is given by
    \begin{align}
        \hat{J}(\! \gamma_l \!) \!\!
            =\!\! 1 \!\!+\! \frac{1}{\gamma_l} \!-\! \frac{1}{4} \! \sqrt{\frac{\pi}{\gamma_l}} \! \exp{ \! \left( \!\! - \frac{\gamma_l}{2} \! \right)} \!\!\! \left[ \! (1 \!+\! \gamma_l^{-1} \!) \! I_0 \!\! \left( \! \frac{\gamma_l}{2} \! \right) \!\!+\! I_1 \!\! \left( \! \frac{\gamma_l}{2} \! \right) \! \right]\!\!.
    \end{align}
\end{rem}

Numerical simulation reveals that the approximation error between the asymptotic equation we use and Eq. (\ref{eqJgamma}) is controlled within 0.07, which gradually converges to 0 with increasing $\gamma$. In wideband scenarios, $\gamma$ is typically large enough so that $\hat{J}(\gamma_l)$ provides an accurate approximation of Eq. (\ref{eqJgamma}) with much simpler calculation.

\section{Numerical Results} \label{secNumRes}
This section shows the simulation results of our proposed wideband channel sensing architecture. We adopt the 3GPP clustered delay line (CDL) channel model, which has 2 scattering clusters, each containing 20 rays, and 40 multipaths in total. The center carrier frequency is set as 3.5 GHz with 30 kHz of subcarrier spacing. The root mean square angular spreads of azimuth angle of departure (AOD), zenith angle of departure (ZOD), AOA and ZOA are $47^\circ, 133^\circ, 98^\circ$ and $82^\circ$, respectively. HIS is composed of four rows and four columns of single-polarized radiation units. The number of hologram samples is fixed to $L = 100$ during one time slot. The estimation error is defined as
\begin{align}
    \text{NMSE} = 10 \log \left\{ \mathbb{E}\left[ \frac{\|\hat{\mathbf{h}} - \mathbf{h} \|_2^2}{\| \mathbf{h} \|_2^2} \right] \right\},
\end{align}
where $\hat{\mathbf{h}}$ and $\mathbf{h}$ are the estimated and exact channels, respectively, and the expectation is taken over time.

\begin{figure}[!t]
    \centering
    \includegraphics[width=\columnwidth]{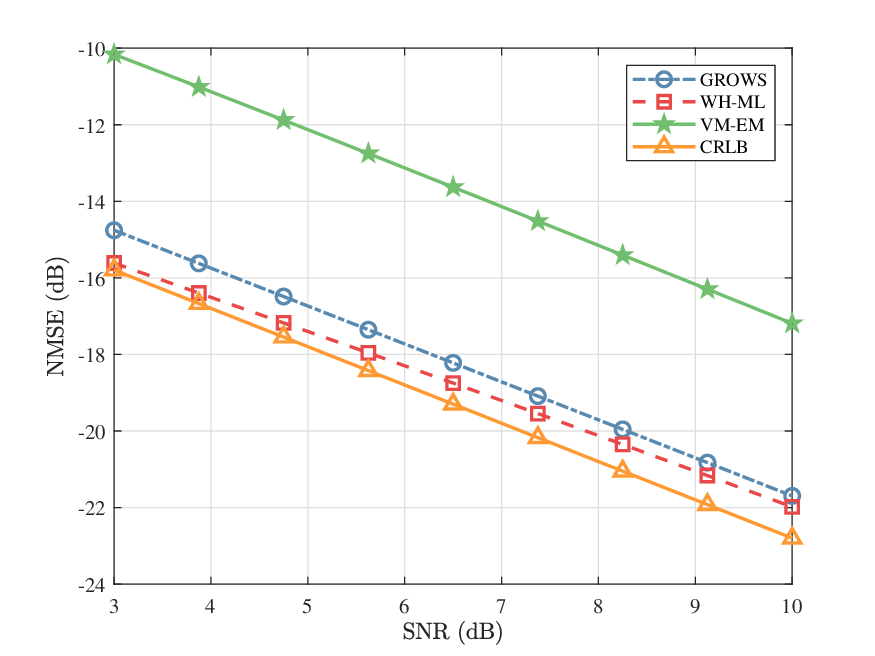}
    \caption{The channel estimation error vs. SNR. The number of RBs is 11.}
    \label{figSim_SNR_Nf10}
\end{figure}

\begin{figure}[!t]
    \centering
    \includegraphics[width=\columnwidth]{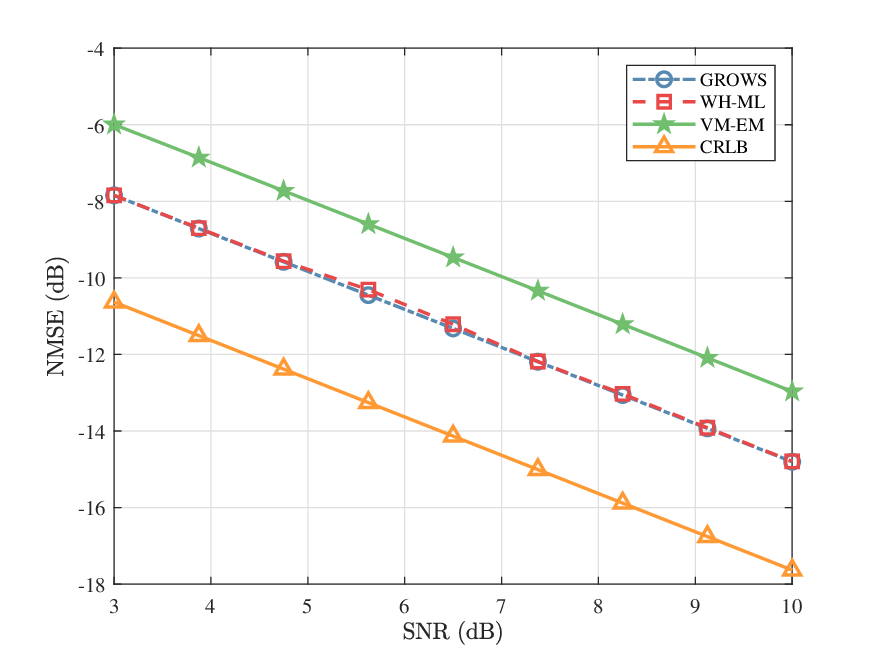}
    \caption{The channel estimation error vs. SNR. The number of RBs is 140.}
    \label{figSim_SNR_Nf35}
\end{figure}

Fig. \ref{figSim_SNR_Nf10} and Fig. \ref{figSim_SNR_Nf35} compare the performance of our proposed GROWS algorithm with WH-ML and von Mises-expectation maximization (VM-EM) method \cite{Zhu:23TIT} in terms of the channel estimation error. The simulation results for a bandwidth of 5 MHz, which contains 11 resource blocks (RBs), are presented in Fig. \ref{figSim_SNR_Nf10}. Meanwhile, Fig. \ref{figSim_SNR_Nf35} illustrates the scenario with 140 RBs, where the bandwidth exceeds 50 MHz. Every single RB contains 12 subcarriers. VM-EM method is an iterative EM algorithm-based phase estimator for chi-square distributions. Since VM-EM method only estimates the phase, we take the phase of $\hat{\mathbf{h}}_{\text{GROWS}}$, i.e. the estimation result of GROWS algorithm, as the initial value for VM-EM method, and then update the channel estimate according to the VM-EM rules. It is shown that our proposed GROWS algorithm provides a closer estimation result to the CRLB with significantly low computational complexity. Due to the complex scenario of wideband holographic channel estimation, the WH-ML estimator only provides a limited performance gain over the GROWS algorithm, which is further reduced as the number of subcarriers increases.

\begin{figure}[!t]
    \centering
    \includegraphics[width=\columnwidth]{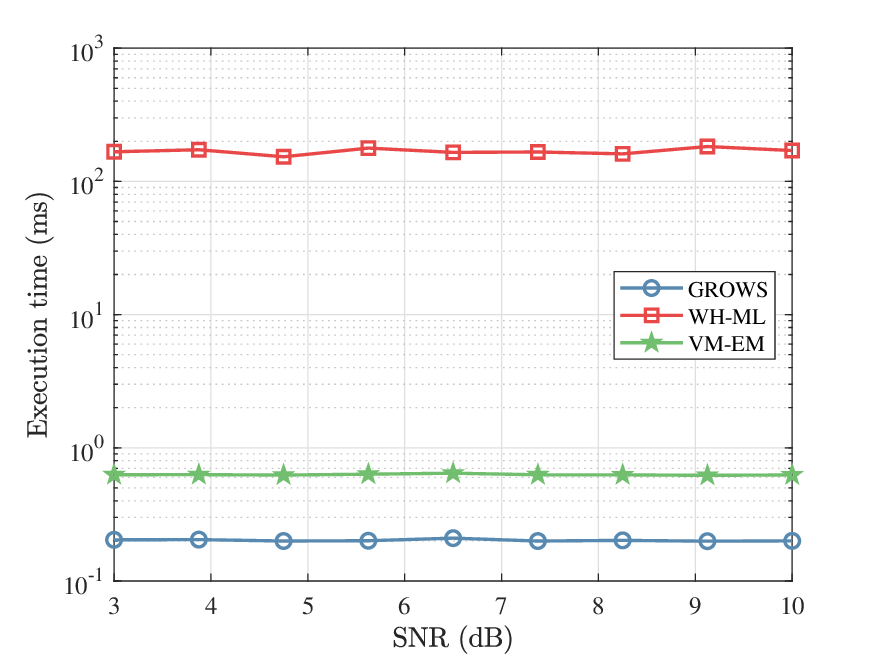}
    \caption{Execution time for different holographic channel estimation methods.}
    \label{figSim_Time}
\end{figure}

Fig. \ref{figSim_Time} presents the execution time of three different holographic channel estimation methods. Each iteration of the WH-ML method incurs high computational cost from calculating the complex Hessian matrix and Bessel functions, which requires complex matrix multiplications and solving equation systems. Conversely, the GROWS method recovers the linear structure via element-wise operations, leading to a computational time nearly three orders of magnitude lower than WH-ML. The EM algorithm was designed to simplify ML computation and ensure rapid convergence, enabling VM-EM to achieve execution time on the same order of magnitude as GROWS. The computational framework of the VM-EM algorithm \cite{Zhu:23TIT} also relies exclusively on element-wise operations and FFT, resulting in a space complexity of $\mathcal{O}(L N_t)$.

\begin{figure}[!t]
    \centering
    \includegraphics[width=\columnwidth]{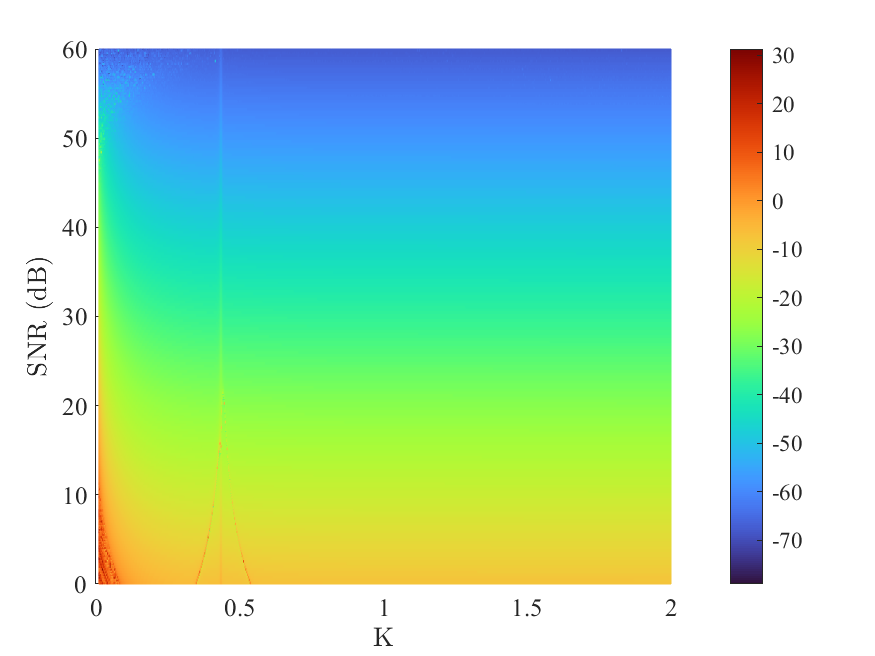}
    \caption{CRLB as a two-variable function of SNR and K.}
    \label{figSim_CRLB}
\end{figure}

Fig. \ref{figSim_CRLB} plots CRLB as a two-variable function of signal-to-noise ratio (SNR) and the power of reference waves. For the purpose of describing the relative relationship between the power of the reference wave and the object wave, we define $K = |E_r(t)|/\max{\{ |E_o(t)| \}}$. According to the analysis we performed in Sec. \ref{secHCE_CRLB}, the CRLB is mainly dependent on the noise power and the reference wave intensity. The estimation accuracy can be dramatically improved for wideband holographic communication scenarios by enhancing the reference wave intensity when $K < 0.2$. In more general cases, the larger SNR contributes more to accurate holographic channel estimation. Note that the CRLB is a function of $\mu_l$, i.e., the mean value of the non-central chi-squared distribution, which is determined by the channel coefficient $h_k$ and reference wave intensity $A_r$. This makes the specific form of the CRLB dependent on the channel structure in the low K-value region. Conversely, in the high K-value region, the reference wave suppresses the structural characteristics of the channel, such that the CRLB exhibits similar properties across different channels.

\begin{figure}[!t]
    \centering
    \includegraphics[width=\columnwidth]{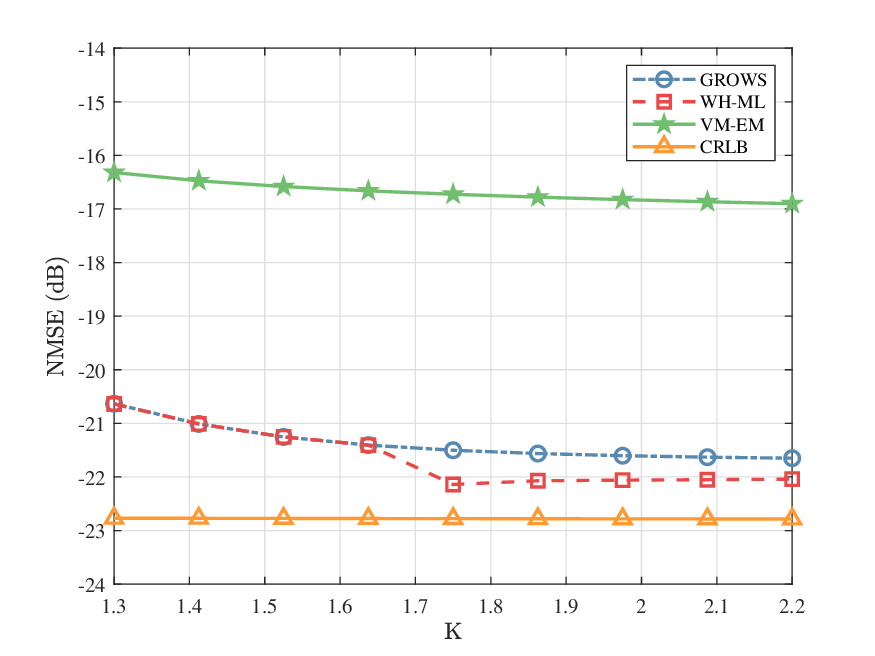}
    \caption{The channel estimation error vs. K, SNR = 10 dB.}
    \label{figSim_K}
\end{figure}

\begin{figure}[!t]
    \centering
    \includegraphics[width=\columnwidth]{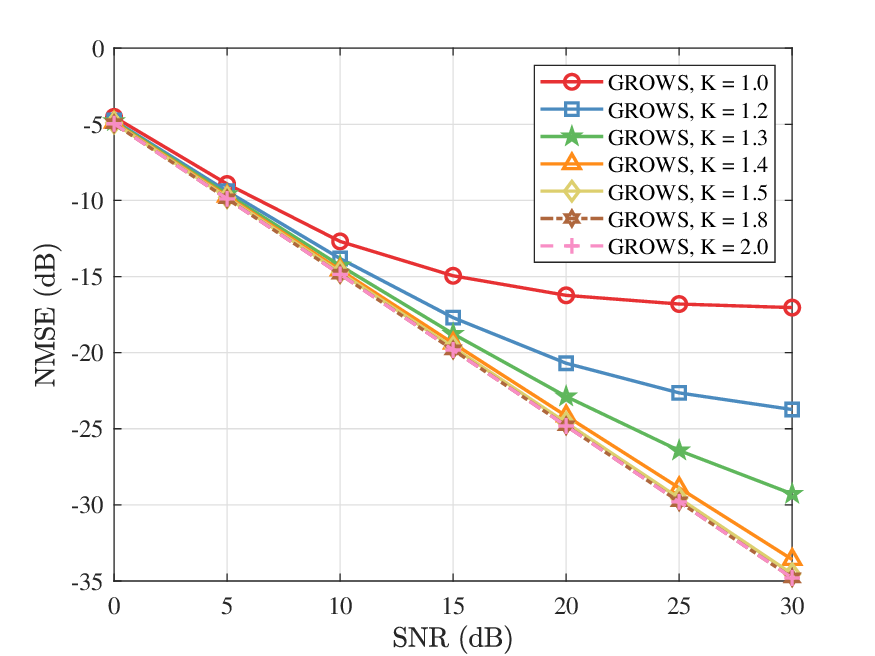}
    \caption{The channel estimation error vs. SNR with different reference wave power.}
    \label{figSim_SNRk}
\end{figure}

The estimation error as a function of the power of reference waves is illustrated in Fig. \ref{figSim_K} and Fig. \ref{figSim_SNRk}. We may observe from Fig. \ref{figSim_K} that increasing reference wave power significantly improves the channel estimation accuracy for given noise power. As we analyzed in Sec. \ref{secHoloComm_HCR}, the process of recovering the object wave in the GROWS algorithm is approximated to a localization problem, besides the reference wave is generated by the HIS, being almost unaffected by the electromagnetic noise in the environment. Therefore, the impact of the perturbation caused by electromagnetic noise on the localization accuracy of the object wave decreases with the enhancement of the reference wave power. Fig. \ref{figSim_SNRk} illustrates the influence of reference wave power on the channel estimation error under different SNR. For high SNR scenarios, the enhancement of estimation accuracy by the reference wave power is considerably intensified, which reflects the localization properties of the holographic channel recovery.

\section{Conclusion}\label{secConcl}
In this paper, we investigated an interference principle-based wideband channel sensing architecture for HIS. The holographic interference principle has been extended to wideband communication scenarios for signal receiving and channel estimation. Based on the properties and physical significance of wideband holograms, we have proposed a holographic channel recovery method for overcoming the spectral overlapping caused by the nonlinear structure of holograms. The maximum likelihood estimation based on the holograms has also been developed. The GROWS algorithm is then proposed as a generalized wideband holographic channel estimation method with much lower complexity. Furthermore, we have derived the CRLB of the wideband holographic channel estimation problem. Simulation results have also verified the effectiveness of our method and the simplification of wideband channel sensing architecture for channel estimation.

\appendices
\section{Calculation of the Hessian Matrix} \label{apxHess}
According to the definition of the Hessian matrix, the second order partial derivatives of $\mathcal{F}(\mathbf{h})$ are required to be evaluated.  The following shows the derivation of the Hessian matrix with $\boldsymbol{\mathcal{H}}_{\mathbf{h}^*, \mathbf{h}} (\mathcal{F}(\mathbf{h}))$ as an example.

By the partial derivative property of the Bessel function \cite{Silverman:72}, it follows that $R'(z) = 1 - R^2(z) - R(z)/z$. Therefore, the following equation is obtained by the chain rule
\begin{align}
    \frac{\partial R(z_l)}{\partial \mathfrak{R}(h_n)}
		=& \frac{\partial R(z_l)}{\partial z_l} \frac{\partial z_l}{\partial \mathfrak{R}(h_n)} \nonumber \\
		=&  \left( 1 - R^2(z_l) - \frac{R(z_l)}{z_l} \right) \frac{2\sqrt{\tilde{E}_I[l]}}{\sigma_{\omega}^2} \frac{\partial |\mu_l|}{\partial \mathfrak{R}(h_n)}. 
\end{align}
According to the law of derivation for real functions, the partial derivative of $\frac{\mu_l^*}{|\mu_l|}$ with respect to $\mathfrak{R}(h_m)$ is given by
\begin{align}
    \frac{\partial}{\partial \mathfrak{R}(h_n)} \left( \frac{\mu_l^*}{|\mu_l|} \right)
		=& \frac{1}{|\mu_l|^2} \left\{ |\mu_l| \frac{\partial \mu_l^*}{\partial \mathfrak{R}(h_n)} - \mu_l^* \frac{\partial |\mu_l|}{\partial \mathfrak{R}(h_n)} \right\} \nonumber \\
		=& \frac{1}{|\mu_l|} \frac{\partial \mu_l^*}{\partial \mathfrak{R}(h_n)} - \frac{\mu_l^*}{|\mu_l|^2} \frac{\partial |\mu_l|}{\partial \mathfrak{R}(h_n)}.
\end{align}
By substituting the two equations above, we may derive
\begin{align}
	&\frac{\partial}{\partial \mathfrak{R}(h_n)} \left( \frac{\mu_l^*}{|\mu_l|} R(z_l) \right) \nonumber \\
	&= \!R(z_l) \frac{\partial}{\partial \mathfrak{R}(h_n)} \left( \frac{\mu_l^*}{|\mu_l|} \right) + \frac{\mu_l^*}{|\mu_l|} \frac{\partial R(z_l)}{\partial \mathfrak{R}(h_n)} \nonumber \\
	&= \!\frac{R(\!z_l\!)}{|\mu_l|} \frac{\partial \mu_l^*}{\partial \mathfrak{R}(h_n)} \!+\! \left( z_l \!-\! z_l R^2(\!z_l\!) \!-\! 2 R(\!z_l\!) \!\right)\! \frac{\mu_l^*}{|\mu_l|^2} \frac{\partial |\mu_l|}{\partial \mathfrak{R}(\!h_n\!)}\!. \label{eqApxHessR}
\end{align}
Similarly, we may obtain the partial derivative of $\frac{\mu_l^*}{|\mu_l|} R(z_l)$ with respect to $\mathfrak{I}(h_m)$ as
\begin{align}
    &\frac{\partial}{\partial \mathfrak{I}(h_n)} \left( \frac{\mu_l^*}{|\mu_l|} R(z_l) \right) \nonumber \\
	&= \frac{R(\!z_l\!)}{|\mu_l|} \frac{\partial \mu_l^*}{\partial \mathfrak{I}(h_n)} \!+\! \left(\! z_l \!-\! z_l R^2(\!z_l\!) \!-\! 2 R(\!z_l\!) \!\right)\! \frac{\mu_l^*}{|\mu_l|^2} \frac{\partial |\mu_l|}{\partial \mathfrak{I}(\!h_n\!)}. \label{eqApxHessI}
\end{align}
Substituting Eq. (\ref{eqApxHessR}) and Eq. (\ref{eqApxHessI}) into the definition of Wirtinger partial derivative, it follows that
\begin{align}
    &\frac{\partial}{\partial h_n} \left( \frac{\mu_l^*}{|\mu_l|} R(z_l) \right) \nonumber \\
	&= \frac{1}{2} \left[ \frac{\partial}{\partial \mathfrak{R}(h_n)} \left( \frac{\mu_l^*}{|\mu_l|} R(z_l) \right) - j\frac{\partial}{\partial \mathfrak{I}(h_n)} \left( \frac{\mu_l^*}{|\mu_l|} R(z_l) \right) \right] \nonumber \\
	&= \frac{R(z_l)}{|\mu_l|} \frac{\partial \mu_l^*}{\partial h_m} + \left( z_l - z_l R^2(z_l) - 2 R(z_l) \right) \frac{\mu_l^*}{|\mu_l|^2} \frac{\partial |\mu_l|}{\partial h_n} \nonumber \\
	&= \left( \frac{z_l}{2} - \frac{z_l}{2} R^2(z_l) - R(z_l) \right) \frac{|\mu_l| \Phi_{l,n}}{\mu_l^2}.
\end{align}
Thus, the second order partial derivative of $\mathcal{F}(\mathbf{h})$ is given by
\begin{align}
    &\frac{\partial^2 \mathcal{F}(\mathbf{h})}{\partial h_m^* \partial h_n} \nonumber \\
    &= \frac{1}{\sigma_{\omega}^2} \! \sum_{l=1}^{L} 2 \sqrt{\tilde{E}_I[l]} \Phi_{l,m}^* \frac{\partial}{\partial h_n} \! \left( \frac{\mu_l^*}{|\mu_l|} R(z_l) \right) \!-\! \frac{1}{\sigma_{\omega}^2} \! \sum_{l=1}^{L} \Phi_{l,m}^* \frac{\partial \mu_l}{\partial h_n} \nonumber \\
    &= \frac{1}{\sigma_{\omega}^4} \sum_{l=1}^{L} \left( 2 \tilde{E}_I[l] - 2 \tilde{E}_I[l] R^2(z_l) - \sigma_{\omega}^2 \right) \Phi_{l,m}^* \Phi_{l,n}.
\end{align}
And the partial Hessian matrix $\boldsymbol{\mathcal{H}}_{\mathbf{h}^*, \mathbf{h}} (\mathcal{F}(\mathbf{h}))$ is expressed as
\begin{align}
    \boldsymbol{\mathcal{H}}_{\mathbf{h}^*, \mathbf{h}} (\mathcal{F}(\mathbf{h}))
		= \frac{\partial^2 \mathcal{F}(\mathbf{h})}{\partial \mathbf{h}^* \partial \mathbf{h}^T}
		= \mathbf{\Phi}^* \mathbf{R}_2(\mathbf{h}) \mathbf{\Phi}^T.
\end{align}
Through consecutive similar procedures, we may obtain
\begin{align}
    \frac{\partial}{\partial h_n} \left( \frac{\mu_l}{|\mu_l|} R(z_l) \right)
		=& \frac{z_l - z_l R^2(z_l)}{2 |\mu_l|} \Phi_{l,n}, \\
	\frac{\partial}{\partial h_n^*} \left( \frac{\mu_l}{|\mu_l|} R(z_l) \right)
		=& \frac{z_l - z_l R^2(z_l) - 2 R(z_l)}{2 |\mu_l|^3} \mu_l^2 \Phi_{l,n}^*, \\
	\frac{\partial}{\partial h_n^*} \left( \frac{\mu_l^*}{|\mu_l|} R(z_l) \right)
		=& \frac{z_l - z_l R^2(z_l)}{2 |\mu_l|} \Phi_{l,n}^*.
\end{align}
The rest of the second order partial derivatives are given by 
\begin{align}
	\frac{\partial^2 \mathcal{F}(\mathbf{h})}{\partial h_m^* \partial h_n^*}
		\!=& \sum_{l=1}^{L} \frac{ z_l^2 - z_l^2 R^2(z_l) - 2 z_l R(z_l)}{2 (\mu_l^*)^2} \Phi_{l,m}^* \Phi_{l,n}^*, \\
	\frac{\partial^2 \mathcal{F}(\mathbf{h})}{\partial h_m \partial h_n}
		\!=& \sum_{l=1}^{L} \frac{ z_l^2 - z_l^2 R^2(z_l) - 2 z_l R(z_l)}{2 \mu_l^2} \Phi_{l,m} \Phi_{l,n}, \\
	\frac{\partial^2 \mathcal{F}(\mathbf{h})}{\partial h_m \partial h_n^*}
		=& \frac{1}{\sigma_{\omega}^4} \!\! \sum_{l=1}^{L} \!\left( 2 E_I\![l] \!\!-\!\! 2 E_I\![l] R^2(\!z_l\!) \!-\! \sigma_{\omega}^2 \right)\! \Phi_{l,m} \! \Phi_{l,n}^*,
\end{align}
which completes the calculation of the Hessian Matrix. $\hfill\IEEEQED$

\section{Proof of Theorem \ref{thmCRLB}} \label{apxThmCRLB}
According to \cite{Ollila:08}, the CRLB of complex parameters is defined as
\begin{align}
    \text{CRLB}(\mathbf{h}) = 
        \begin{bmatrix} 
			\boldsymbol{\mathcal{I}}_{\mathbf{h}}   & \boldsymbol{\mathcal{P}}_{\mathbf{h}} \\
			\boldsymbol{\mathcal{P}}_{\mathbf{h}}^* & \boldsymbol{\mathcal{I}}_{\mathbf{h}}^*
		\end{bmatrix}^{-1},
\end{align}
where the complex information matrix $\boldsymbol{\mathcal{I}}_{\mathbf{h}} \in \mathbb{C}^{N_f \times N_f}$ and the pseudo-information matrix $\boldsymbol{\mathcal{P}}_{\mathbf{h}} \in \mathbb{C}^{N_f \times N_f}$ are given by
\begin{align}
    \boldsymbol{\mathcal{I}}_{\mathbf{h}} 
		=& \mathbb{E} [\nabla_{\mathbf{h}^*} \mathcal{F}(\mathbf{h}) \{ \nabla_{\mathbf{h}^*} \mathcal{F}(\mathbf{h}) \}^H], \\
	\boldsymbol{\mathcal{P}}_{\mathbf{h}} 
		=& \mathbb{E} [\nabla_{\mathbf{h}^*} \mathcal{F}(\mathbf{h}) \{ \nabla_{\mathbf{h}^*} \mathcal{F}(\mathbf{h}) \}^T].
\end{align}
According to Eq. (\ref{eqGradm}), we may derive
\begin{align}
    &\frac{\partial \mathcal{F}(\mathbf{h})}{\partial h_m^*} \cdot \frac{\partial \mathcal{F}(\mathbf{h})}{\partial h_n} \nonumber \\
    &= \!\! \frac{1}{\sigma_{\omega}^4} \!\! \sum_{l_1=1}^{L} \!\! \sum_{l_2=1}^{L} \!\! \left(\!\! \frac{2 \sqrt{\tilde{E}_I[l_1]}}{|\mu_{l_1}|} \!\! R(z_{l_1}) \!-\!\! 1 \!\! \right) \!\!\!\! \left(\!\! \frac{2 \sqrt{\tilde{E}_I[l_2]}}{|\mu_{l_2}|} \!\! R(z_{l_2}) \!-\!\! 1 \!\! \right)\!\! \varphi_{l_1,l_2} \nonumber \\
    &=\!\! \frac{1}{\sigma_{\omega}^4} \!\! \sum_{l_1=1}^{L} \!\! \sum_{\substack{l_2=1 \\ l_2 \neq l_1}}^{L} \!\! \left( \!\! \frac{2 \sqrt{\tilde{E}_I[l_1]}}{|\mu_{l_1}|} \!\! R(z_{l_1}) \!-\!\! 1 \!\! \right) \!\!\!\! \left( \!\! \frac{2 \sqrt{\tilde{E}_I[l_2]}}{|\mu_{l_2}|} \!\! R(z_{l_2}) \!-\!\! 1 \!\! \right)\!\! \varphi_{l_1,l_2} \nonumber \\
    &+ \frac{1}{\sigma_{\omega}^4} \sum_{l=1}^{L} \!\left( \frac{4 \tilde{E}_I[l]}{|\mu_l|^2} R^2(z_l) - 4 \frac{\sqrt{\tilde{E}_I[l]}}{|\mu_l|} R(z_l) + 1 \right) \!\! \varphi_{l,l},
\end{align}
where $\varphi_{l_1,l_2} = \mu_{l_1} \! \mu_{l_2}^* \! \Phi_{l_l,m}^* \! \Phi_{l_2,n}$. It is evident from the above equation that $\mathbb{E} \left[\frac{\sqrt{\tilde{E}_I[l]}}{|\mu_l|} R(z_l) \right]$ and $\mathbb{E} \left[\frac{\tilde{E}_I[l]}{|\mu_l|^2} R^2(z_l) \right]$ are required for the calculation of $\mathbb{E} \left[\frac{\partial \mathcal{F}(\mathbf{h})}{\partial h_m^*} \cdot \frac{\partial \mathcal{F}(\mathbf{h})}{\partial h_n} \right]$. Based on the derivation in \cite{Zhu:23TIT}, $\mathbb{E} \left[\frac{\tilde{E}_I[l]}{|\mu_l|^2} R^2(z_l) \right]$ is obtained by the definition of expectation as
\begin{align}
    &\mathbb{E} \left[\frac{\tilde{E}_I[l]}{|\mu_l|^2} R^2(z_l) \right] \nonumber \\
    &= \! \int_{0}^{+\infty} \!\!\!\!\!\! \frac{x}{\sigma_{\omega}^2 |\mu_l|^2} \! R^2 \!\! \left( 2\frac{\sqrt{x}|\mu_l|}{\sigma_{\omega}^2} \right) \! \exp \! \left(-\frac{x}{\sigma_{\omega}^2}\right) \! I_0 \! \left(\frac{2\sqrt{x}|\mu_l|}{\sigma_{\omega}^2}\right) \! dx \nonumber \\
	&= \int_{0}^{+\infty} \!\!\!\!\!\!\!\! \gamma_l t e^{-\gamma_l(1+t)} \! I_0 \!\left( 2 \gamma_l \sqrt{t} \right)\! R^2 \!\left( 2 \gamma_l \sqrt{t} \right) dt 
	\overset{\Delta}{=} J(\gamma_l),
\end{align}
where $x = |\mu_l|^2 t$ and $\gamma_l = \frac{|\mu_l|^2}{\sigma_{\omega}^2}$. It follows from the property of expectation that 
\begin{align}
    &\int_{0}^{+\infty} \frac{1}{\sigma_{\omega}^2} \exp\left(-\frac{x}{\sigma_{\omega}^2}\right) I_0\left(\frac{2\sqrt{x}|\mu_l|}{\sigma_{\omega}^2}\right) dx \nonumber \\
	&= \exp{\left( \frac{|\mu_l|^2}{\sigma_{\omega}^2} \right)} \!\! \int_{0}^{+\infty} \!\! \frac{1}{\sigma_{\omega}^2} \! \exp \! \left(-\frac{x \!+\! |\mu_l|^2}{\sigma_{\omega}^2}\right) \! I_0 \! \left(\frac{2\sqrt{x}|\mu_l|}{\sigma_{\omega}^2}\right) dx \nonumber \\
	&= \exp{\left( \frac{|\mu_l|^2}{\sigma_{\omega}^2} \right)}.
\end{align}
Therefore, $\mathbb{E} \left[\frac{\sqrt{\tilde{E}_I[l]}}{|\mu_l|} R(z_l) \right]$ is given by
\begin{align}
    &\mathbb{E} \left[\frac{\sqrt{\tilde{E}_I[l]}}{|\mu_l|} R(z_l) \right] \nonumber \\
    &= \int_{0}^{+\infty} \frac{1}{\sigma_{\omega}^2} \exp\left(-\frac{x+|\mu_l|^2}{\sigma_{\omega}^2}\right) I_1\left(\frac{2\sqrt{x}|\mu_l|}{\sigma_{\omega}^2}\right) \frac{\sqrt{x}}{|\mu_l|} dx \nonumber \\
	&= \!\! \frac{\partial}{\partial |\mu_l|^2} \!\!\! \int_{0}^{+\infty} \!\!\!\! \frac{1}{\sigma_{\omega}^2} \! \exp \!\left( \!\! -  \frac{x}{\sigma_{\omega}^2} \!\!\right) \!\! I_0 \!\! \left( \! \frac{2\sqrt{x}|\mu_l|}{\sigma_{\omega}^2} \! \right) \!\! dx \! \times \! \sigma_{\omega}^2 \exp{ \!\left( \!\! -\frac{|\mu_l|^2}{\sigma_{\omega}^2} \!\right)} \nonumber \\
    &= \frac{\partial}{\partial |\mu_l|^2} \exp{\left( \frac{|\mu_l|^2}{\sigma_{\omega}^2} \right)} \times \sigma_{\omega}^2 \exp{\left(-\frac{|\mu_l|^2}{\sigma_{\omega}^2}\right)} = 1.
\end{align}
Since $\tilde{E}_I[l]$ is independent of each other, we may obtain
\begin{align}
    &\mathbb{E} \left[\frac{\partial \mathcal{F}(\mathbf{h})}{\partial h_m^*} \cdot \frac{\partial \mathcal{F}(\mathbf{h})}{\partial h_n} \right] \nonumber \\
    &= \!\! \frac{1}{\sigma_{\omega}^4} \!\! \sum_{l_1=1}^{L} \!\! \sum_{\substack{l_2=1 \\ l_2 \neq l_1}}^{L} \!\! \mu_{l_1} \! \mu_{l_2}^* \! \Phi_{l_1,m}^* \! \Phi_{l_2,n}
		\!\!+\!\! \frac{1}{\sigma_{\omega}^4} \!\!\sum_{l=1}^{L} \!\! \left( 4 J(\gamma_l) \!\!-\!\! 3 \right) \! |\mu_l|^2 \Phi_{l,m}^* \! \Phi_{l,n} \nonumber \\
	&= \!\! \frac{1}{\sigma_{\omega}^4} \!\! \sum_{l_1=1}^{L} \!\! \mu_{l_1} \! \Phi_{l_l,m}^* \!\! \sum_{l_2=1}^{L} \!\! \mu_{l_2}^* \! \Phi_{l_2,n}
		\!+\! \frac{4}{\sigma_{\omega}^4} \!\! \sum_{l=1}^{L} \!\! \left(J(\gamma_l) \!-\! 1 \right) \!\! |\mu_l|^2 \Phi_{l,m}^* \! \Phi_{l,n}. \label{eqIhmn}
\end{align}
Similarly, the expectation of $\frac{\partial \mathcal{F}(\mathbf{h})}{\partial h_m^*} \cdot \frac{\partial \mathcal{F}(\mathbf{h})}{\partial h_n^*}$ is given by
\begin{align}
    \mathbb{E} \! \left[ \! \frac{\partial \mathcal{F}(\mathbf{h})}{\partial h_m^*} \!\cdot\! \frac{\partial \mathcal{F}(\mathbf{h})}{\partial h_n^*} \! \right] \!
    =&  \frac{1}{\sigma_{\omega}^4} \!\! \sum_{l_1=1}^{L} \!\! \mu_{l_1} \Phi_{l_l,m}^* \!\! \sum_{l_2=1}^{L} \!\! \mu_{l_2} \Phi_{l_2,n}^* \nonumber \\
		&+\! \frac{4}{\sigma_{\omega}^4} \!\! \sum_{l=1}^{L} \!\! \left(J(\gamma_l) \!-\! 1 \right) \! \mu_l^2 \Phi_{l,m}^* \! \Phi_{l,n}^*. \label{eqPhmn}
\end{align}

Substituting Eq. (\ref{eqIhmn}) and Eq. (\ref{eqPhmn}) into the definition of information matrix, we obtain
\begin{align}
    \boldsymbol{\mathcal{I}}_{\mathbf{h}} \!\!
		=& \mathbb{E} [\nabla_{\mathbf{h}^*} \mathcal{F}(\mathbf{h}) \{ \nabla_{\mathbf{h}^*} \mathcal{F}(\mathbf{h}) \}^H]
		\!=\!\!  \frac{1}{\sigma_{\omega}^4} \! \mathbf{\Phi}^* \!\! \left(\! \boldsymbol{\mu} \boldsymbol{\mu}^H \!\!+\! \boldsymbol{\mathcal{J}}_{\!\mathcal{I}} \!\right)\!\! \mathbf{\Phi}^T\!\!, \\
	\boldsymbol{\mathcal{P}}_{\mathbf{h}} \!\!
		=& \mathbb{E} [\nabla_{\mathbf{h}^*} \mathcal{F}(\mathbf{h}) \{ \nabla_{\mathbf{h}^*} \mathcal{F}(\mathbf{h}) \}^T]
		\!=\!\!  \frac{1}{\sigma_{\omega}^4} \! \mathbf{\Phi}^* \!\! \left(\! \boldsymbol{\mu} \boldsymbol{\mu}^T \!\!+\! \boldsymbol{\mathcal{J}}_{\!\mathcal{P}} \!\right)\!\! \mathbf{\Phi}^H\!\!,
\end{align}
where $\boldsymbol{\mathcal{J}}_{\mathcal{I}} = \text{diag}\left\{ 4 \left( J(\gamma_l)-1 \right) |\mu_l|^2 \right\}$ and $\boldsymbol{\mathcal{J}}_{\mathcal{P}} = \text{diag}\left\{ 4 \left( J(\gamma_l)-1 \right) \mu_l^2 \right\}$. Utilizing the result for the inverse of a partitioned matrix, the CRLB of $\mathbf{h}$ is given by
\begin{align}
    \text{CRLB}(\mathbf{h}) \!=\!\! 
        \begin{bmatrix} 
			\boldsymbol{\mathcal{I}}_{\mathbf{h}}   \!&\! \boldsymbol{\mathcal{P}}_{\mathbf{h}} \\
			\boldsymbol{\mathcal{P}}_{\mathbf{h}}^* \!&\! \boldsymbol{\mathcal{I}}_{\mathbf{h}}^*
		\end{bmatrix}^{-1} \!\!=\!\! 
        \begin{bmatrix}
            \boldsymbol{\mathcal{R}}_{\mathbf{h}}^{-1} \!&\! -\boldsymbol{\mathcal{R}}_{\mathbf{h}}^{-1} \boldsymbol{\mathcal{Q}}_{\mathbf{h}} \\
			-\boldsymbol{\mathcal{Q}}_{\mathbf{h}}^{H} \boldsymbol{\mathcal{R}}_{\mathbf{h}}^{-1} \!&\! (\boldsymbol{\mathcal{R}}_{\mathbf{h}}^{-1})^*
        \end{bmatrix}\!,
\end{align}
where $\boldsymbol{\mathcal{R}}_{\mathbf{h}} = \boldsymbol{\mathcal{I}}_{\mathbf{h}} - \boldsymbol{\mathcal{P}}_{\mathbf{h}} (\boldsymbol{\mathcal{I}}_{\mathbf{h}}^{-1})^* \boldsymbol{\mathcal{P}}_{\mathbf{h}}^*$ and $\boldsymbol{\mathcal{Q}}_{\mathbf{h}} = \boldsymbol{\mathcal{P}}_{\mathbf{h}} (\boldsymbol{\mathcal{I}}_{\mathbf{h}}^{-1})^*$. Thus, Theorem \ref{thmCRLB} is proved. $\hfill\IEEEQED$

\bibliographystyle{IEEEtran}
\bibliography{ref}
 
\vfill

\end{document}